\documentclass[twocolumn,tighten,times]{aastex63}
\usepackage{graphicx}
\usepackage{amssymb}
\usepackage{amsbsy}
\usepackage{amsmath}
\usepackage{mathrsfs}
\usepackage{color}
\usepackage{hyperref}
\usepackage{bm}

\received{2022 July 25}
\revised{2022 November 11}
\accepted{2022 November 25}
\published{2023 January 23}
\shorttitle{Spectroastrometric Reverberation Mapping}
\shortauthors{Li \& Wang}

\begin{document}

\title{\bf \large Spectroastrometric Reverberation Mapping of Broad-line Regions}
\author[0000-0001-5841-9179]{Yan-Rong Li}
\affiliation{Key Laboratory for Particle Astrophysics, Institute of High Energy Physics, Chinese Academy of Sciences, 19B Yuquan Road, Beijing 100049, China;
\href{mailto:liyanrong@mail.ihep.ac.cn}{liyanrong@mail.ihep.ac.cn}, \href{mailto:wangjm@mail.ihep.ac.cn}{wangjm@mail.ihep.ac.cn}}

\author[0000-0001-7617-4232]{Jian-Min Wang}
\affiliation{Key Laboratory for Particle Astrophysics, Institute of High Energy Physics, Chinese Academy of Sciences, 19B Yuquan Road, Beijing 100049, China; \href{mailto:liyanrong@mail.ihep.ac.cn}{liyanrong@mail.ihep.ac.cn}, \href{mailto:wangjm@mail.ihep.ac.cn}{wangjm@mail.ihep.ac.cn}}
\affil{University of Chinese Academy of Sciences, 19A Yuquan Road, Beijing 100049,  China}
\affil{National Astronomical Observatories of China, Chinese Academy of Sciences, 20A Datun Road, Beijing 100020, China}

\begin{abstract}
Spectroastrometry measures source astrometry as a function of wavelength/velocity. Reverberations of spectroastrometric 
signals naturally arise in broad-line regions (BLRs) of active galactic nuclei (AGNs) as a result of the continuum
variations that drive responses of the broad emission lines with time delays. Such signals provide a new diagnostic for
mapping BLR kinematics and geometry, complementary to the traditional intensity reverberation mapping (RM) technique.
We present the generic mathematical formalism for spectroastrometric RM and show that under
realistic parameters of a phenomenological BLR model, the spectroastrometric reverberation signals vary on a level of several to tens of
microarcseconds, depending on the BLR size, continuum variability, and angular-size distance.
We also derive the analytical expressions of spectroastrometric RM for an inclined ring-like BLR.
We develop a Bayesian framework with a sophisticated Monte Carlo sampling
technique to analyze spectroastrometric data and infer the BLR properties, including the central black hole mass and angular-size distance.
We demonstrate the potential of spectroastrometric RM in spatially resolving BLR kinematics and geometry through a suite of simulation tests.
The application to realistic observation data of 3C~273 obtains tentative, but enlightening results, reinforcing the practical feasibility of conducting
spectroastrometric RM experiments on bright AGNs with the operating Very Large Telescope Interferometer
as well as possibly with the planned next-generation 30 m class telescopes.
\end{abstract}
\keywords{Astrometry (84); Reverberation mapping (2019); Supermassive black holes (1663); Active galactic nuclei (16)}

\section{Introduction}
The so-called broad-line regions (BLRs) of active galactic nuclei (AGNs), responsible for prominent broad emission lines in AGN electromagnetic spectra,
are of central importance to the unveiling of gaseous environments surrounding the central supermassive black holes (SMBHs) in general
and to the direct measurement of SMBH masses in particular. The characteristic sizes of BLRs range from light-days to light-months, yet too compact
at cosmic distances to be spatially resolved by existing instruments. The reverberation mapping (RM) technique stemming from
the early works of \cite{Bahcall1972} and \cite{Blandford1982} provides an effective tool to resolve BLRs by swapping spatial resolution
for time resolution (\citealt{Peterson1993}). Its basic ideas are straightforward: the BLR reprocesses the incident ionizing continuum emitted by the
accretion disk into broad-line emissions with time delays due to the light-travel time from the accretion disk to the BLR. Different parts of
the BLR have different time delays and line-of-sight (LOS) velocities; therefore, analyzing reverberation properties of the broad emission line
with respect to the continuum variations deliver information about the BLR geometry and kinematics. Such information is encoded in the two-dimensional
transfer function in velocity and time-delay plane (also called the velocity-delay map), which can be inferred from spectroscopic monitoring
data. However, strictly speaking, RM probes the BLR structure along two dimensions, namely, the time delay and LOS velocity dimensions. On the isodelay and/or isovelocity surfaces, the BLR structures are indeed degenerated from the viewpoint of RM. In this sense, RM bears its own limitations in probing the full dimensions of the BLR structures and further improvements and/or alternative approaches are still warranted for a more thorough understanding of BLR geometry and kinematics.

The spectroastrometry (SA) technique provides such an alternative in the sense that SA probes new dimensions perpendicular to the LOS (\citealt{Beckers1982, Bailey1998}). In particular, measuring a source's SA, namely, the photocenter as a function of wavelength, can achieve a higher positioning accuracy than the angular resolution of the image by a factor of $\sim1/\sqrt{dN_{\rm ph}/d\nu}$, where $dN_{\rm ph}/d\nu$ is the number of photons received in a spectral pixel bin (\citealt{Beckers1982, Bailey1998}). This means that the SA technique can deliver spatial information about a source once it is sufficiently bright.
Indeed, the SA technique has long seen application in radio and millimeter observations, where making the so-called velocity channel maps is a quite standard procedure, from which source positions can be routinely derived. In the optical/infared, however, early full-fledged applications came around the turn of the 21st century and were mainly concentrated on detecting close binary stars  (e.g., \citealt{Bailey1998b, Baines2004, Porter2004}) and spatially resolving circumstellar environments and stellar surface structures (e.g., \citealt{Takami2001, Takami2003, Whelan2004}) at the level of milliarcseconds. Subsequently, \cite{Gnerucci2010} explored the potential
of using the SA of narrow emission lines from the rotating gaseous disk in galactic nuclei to constrain the kinematics of the disk as well as to measure the mass of the central SMBH. They soon after applied this method to two nearby bright galaxies (\citealt{Gnerucci2011, Gnerucci2013}).

Because of the compact sizes of BLRs, until recently, applications to BLRs had been made possible with the successful observation of the infrared Pa$\alpha$ line of 3C 273 by the GRAVITY instrument on board the Very Large Telescope Interferometer (VLTI; \citealt{Gravity2017}). GRAVITY can achieve an angular astrometric resolution down to $\sim$10~$\mu$as (\citealt{Gravity2018}), sufficient to resolve the BLRs of bright AGNs. This achievement, together with the subsequent successful observations of two other AGNs (\citealt{Gravity2020, Gravity2021}), ushered in a new pathway towards BLR physics.

As in RM, arising from BLR reprocesses, the SA of the BLR also reverberates to the continuum variations.
Because of different time delays and LOS velocities at different BLR parts, the reverberation of the spectroastrometric signals with respect
to the continuum can be used to map BLR geometry and kinematics, complementary to the traditional RM. Such an idea had previously been outlined by \cite{Shen2012}, which numerically showcased {\it astrometric} signals of BLR models with simple geometry. In this work, we further carry forward this idea by
developing a generic mathematical framework for spectroastrometric RM and deriving all invoked formulae. As such, we can calculate the spectroastrometric RM signals given an arbitrary BLR model as well as construct a Bayesian approach to explore the BLR model parameters.

Another remarkable capability of spectroastrometric RM is that it can directly measure the geometric distance of the source  (see also \citealt{Wang2020}). This is because SA observations yield both the intensity and photocenter of the broad emission line with wavelength. The intensity reverberation provides information on the physical size of the BLR as in the traditional RM, while the photocenter reverberation provides information on the angular size of the BLR. A combination of these two lines of information naturally constitutes an elegant probe of the geometric distance using AGNs, which therefore has great potential for cosmology (\citealt{Elvis2002}). Previously, \cite{Wang2020} made the first effort in this direction. They integrated the SA observations of the infrared Pa$\alpha$ line and RM observations of the optical H$\beta$ line in 3C~273 and inferred the angular-size distance to 3C~273 (see also \citealt{Li2022}). A subsequent work from the \cite{Gravity2021b} made an application to another nearby bright AGN NGC 3783. Those applications, however, all neglected the SA reverberation and resorted to combining observations of different emission lines, which might correspond to BLRs different in sizes and/or kinematics (see the discussion in \citealt{Li2022}) so some systematics will likely arise. The spectroastrometric RM proposed in this work is more straightforward and surmounts those issues as it measures a single emission line.

This paper is organized as follows. Section~\ref{sec_sarm} develops the generic framework for BLR spectroastrometric RM, including the basic equations, realistic implementation, and a cross-correlation analysis. Section~\ref{sec_example} presents illustrative examples in which there exists simplified or analytical expressions for spectroastrometric RM and Section~\ref{sec_generic} presents spectroastrometric RM signals for a generic BLR model. In Section~\ref{sec_bayes}, we construct a Bayesian approach to infer BLR model parameters and demonstrate the validity of our approach through a suite of simulation tests. With this approach, we also study the dependence of the Bayesian inferences on the sampling rate and measurement errors of simulated SA data.  In Section~\ref{sec_3c273}, we then apply our approach to the observations of 3C 273 and show the tentatively obtained black hole mass and angular-size distance of 3C 273. A discussion and conclusions are given in Sections~\ref{sec_dis} and \ref{sec_conclusion}, respectively.

Throughout the paper, we use {\it intensity RM} to  refer to the traditional RM (only involving line intensity) so as to distinguish it from {\it spectroastrometric RM}.

\section{Spectroastrometric RM}\label{sec_sarm}
\subsection{Basic Equations}
The emission lines stemming from BLRs respond to the continuum variations with time delays. The spatially extended distribution
of BLR gases leads to a distribution of response amplitude over time delays and velocities. At time $t$, the emission line flux 
is given by 
\begin{eqnarray}
F_{\rm BLR}(v, t) &=& \iiint \epsilon(\bm{r}) F_{\rm c}(t-\tau) f(\bm{r}, \bm{w})
   \delta\left(v+\bm{w}\cdot\bm{n}\right)\nonumber\\
   &&\times\delta\left(\tau-\frac{r-\bm{r}\cdot\bm{n}}{c}\right)
   d\bm{r}d\bm{w}d\tau,
\label{eqn_rm_full}
\end{eqnarray}
where $F_{\rm c}$ is the continuum flux, $\delta(x)$ is the Dirac $\delta$ function, $\bm{n}$ denotes the unit vector of the LOS (which
points from the BLR to the observer), 
$\epsilon(\bm{r})$ denotes the spatial distribution of the response coefficient,
$f(\bm{r}, \bm{w})$ denotes the distribution of velocity $\bm{w}$ at the position $\bm{r}$, and
$v$ is the LOS velocity
related to the observed wavelength as 
\begin{equation}
\lambda = \lambda_0 \left(1 + \frac{v}{c}\right),
\end{equation}
where $\lambda_0$ is the rest-frame wavelength of the emission line under consideration.
By defining an intensity transfer function as
\begin{eqnarray}
\Psi(v, \tau) &=& \iint \epsilon(\bm{r}) f(\bm{r}, \bm{w})
   \delta\left(v+\bm{w}\cdot\bm{n}\right)\nonumber\\
   &&\times\delta\left(\tau-\frac{r-\bm{r}\cdot\bm{n}}{c}\right)
   d\bm{r}d\bm{w},
\label{eqn_stf}
\end{eqnarray}
Equation~(\ref{eqn_rm_full}) can be simplified into (e.g., \citealt{Blandford1982, Peterson1993})
\begin{equation}
F_{\rm BLR}(v, t) = \int \Psi(v, \tau) F_{\rm c}(t-\tau) d\tau.
\label{eqn_rm}
\end{equation}
The intensity transfer function $\Psi(v, \tau)$ is determined by the emissivity and velocity distributions of BLR gases (\citealt{Blandford1982}).
Here, we assume that the BLR remains dynamically stable during the period of observations so that
the time dependence of $\epsilon(\bm{r})$ and $f(\bm{r}, \bm{w})$ is negligible.
This assumption is reasonable provided the period of observations is shorter than the dynamical time 
of BLRs. Integrating Equation~(\ref{eqn_rm}) over velocity yields the integrated line flux 
\begin{equation}
\tilde F_{\rm BLR}(t) = \int \tilde{\Psi}(\tau) F_{\rm c}(t-\tau) d\tau,
\end{equation}
where $\tilde\Psi(\tau)$ is the velocity integral of the intensity transfer function, i.e.,
\begin{equation}
\tilde\Psi(\tau) = \int \Psi(v, \tau)dv.
\end{equation}
Hereafter, we use the tilde symbol over a variable to denote its velocity integral.

Due to the responses of BLRs to continuum variations, the SA may also vary with time, namely,
giving rise to spectroastrometric reverberations. We will show below that SA can be measured
by either a spectrometer or an interferometer.
Similar to Equation~(\ref{eqn_rm_full}), we define the moment of the BLR photons as
\begin{eqnarray}
\bm{M}_{\rm BLR}(v, t) &=&
\iiint \bm{r}_\perp\epsilon(\bm{r})F_{\rm c}(t-\tau) f(\bm{r}, \bm{w}) \nonumber\\
 && \hspace*{-1.2cm}\times
   \delta\left(v+\bm{w}\cdot\bm{n}\right)
   \delta\left(\tau-\frac{r-\bm{r}\cdot\bm{n}}{c}\right)
   d\bm{r}d\bm{w}d\tau,
\label{eqn_xblr}
\end{eqnarray}
where 
\begin{equation}
\bm{r}_\perp = \bm{r} - (\bm{r}\cdot\bm{n})\bm{n},
\end{equation}
extracts the component of $\bm{r}$ perpendicular to $\bm{n}$, namely,  the projection of $\bm{r}$ onto the sky.
It is worth stressing that the moment defined above is not normalized by the source's total flux, somehow different from the normal convention.
As shown below, a such defined moment can be expressed in the framework of RM and facilitates theoretical analysis and model calculations.
Similarly, by defining a spectroastrometric transfer function as 
\begin{eqnarray}
\bm{\varPi}(v, \tau) &=& \iint  \bm{r}_\perp\epsilon(\bm{r}) f(\bm{r}, \bm{w})\nonumber\\
   &&\times\delta\left(v+\bm{w}\cdot\bm{n}\right)\delta\left(\tau-\frac{r-\bm{r}\cdot\bm{n}}{c}\right)
   d\bm{r}d\bm{w},
\label{eqn_satf}
\end{eqnarray}
Equation~(\ref{eqn_xblr}) can be simplified into 
\begin{equation}
\bm{M}_{\rm BLR}(v, t) = \int \bm{\varPi}(v, \tau) F_{\rm c}(t-\tau) d\tau.
\label{eqn_sarm}
\end{equation}
By noting that Equations~(\ref{eqn_rm}) and (\ref{eqn_sarm}) have the same integral form, we therefore call 
such astrometric responses of BLR emissions to continuum variations
as spectroastrometric RM.
With the above defined moment, the photocenter of the BLR is calculated as
\begin{equation}
\bm{\varTheta}_{\rm BLR}(v, t) = \frac{\bm{M}_{\rm BLR}(v, t)}{F_{\rm BLR}(v, t)}.
\label{eqn_photon}
\end{equation}

Integrating Equation~(\ref{eqn_sarm}) over velocity $v$ results in 
\begin{equation}
\bm{\tilde M}_{\rm BLR}(t) = \int \bm{\tilde\varPi}(\tau) F_{\rm c}(t-\tau) d\tau,
\end{equation}
where $\bm{\tilde\varPi}(\tau)$ is the velocity integral of the spectroastrometric transfer function, i.e.,
\begin{equation}
\bm{\tilde\varPi}(\tau) = \int \bm{\varPi}(v, \tau)dv.
\end{equation}
Meanwhile, we define the velocity integral of the photocenter $\bm{\varTheta}(v, t)$ weighted by the line profile as
\begin{equation}
\bm{\tilde\varTheta}_{\rm BLR}(t) = \int \frac{F_{\rm BLR}(v, t)}{\tilde F_{\rm BLR}(t)} \bm{\varTheta}_{\rm BLR}(v, t) dv.
\end{equation}
As a result, we obtain the relation among the photocenter, moment, and line flux
\begin{equation}
\bm{\tilde\varTheta}_{\rm BLR}(t) = \frac{\bm{\tilde M}_{\rm BLR}(t)}{\tilde F_{\rm BLR}(t)}.
\end{equation}

We can also calculate the delay integral of $\Psi(v, t)$ and $\bm{\varPi}(v, t)$ as
\begin{equation}
	\hat\Psi(v) = \int \Psi(v, \tau)d\tau,~~~
	\bm{\hat\varPi}(v) = \int  \bm{\varPi}(v, \tau)d\tau,
\end{equation}
where hereafter we use the hat symbol over a variable to denote its delay integral.
It is easy to prove that the mean line profile $\bar F_{\rm BLR}(v)$ is proportional to $\hat\Psi(v)$ and the mean moment
$\bm{\bar M}_{\rm BLR}(v)$ is proportional to $\bm{\hat\varPi}(v)$. Here, $\bar F_{\rm BLR}(v)$ and $\bm{\bar M}_{\rm BLR}(v)$ are defined as
time averages over a time duration $T$,
\begin{eqnarray}
\bar{F}_{\rm BLR}(v)& = &\frac{1}{T}\int_T F_{\rm BLR}(v, t) d t = \hat\Psi(v) {\bar F}_{\rm c}, \label{eqn_mean_fline}\\
\bm{\bar M}_{\rm BLR}(v) &=& \frac{1}{T}\int_T \bm{M}_{\rm BLR}(v, t) dt = \bm{\hat\varPi}(v){\bar F}_{\rm c}, \label{eqn_mean_mom}
\end{eqnarray}
where ${\bar F}_{\rm c}$ is the mean continuum flux over the time duration. Similarly, if defining the mean photocenter as the time average
of $\bm{\varTheta}(v, t)$ weighted by the line profile, namely,
\begin{equation}
\bm{\bar \varTheta}(v) = \frac{1}{T} \int_T \frac{F_{\rm BLR}(v, t)}{\bar{F}_{\rm BLR}(v)} \bm{\varTheta}(v, t) dt,
\end{equation}
we have the relation
\begin{equation}
\bm{\bar \varTheta}(v) = \frac{\bm{\bar M}_{\rm BLR}(v)}{\bar{F}_{\rm BLR}(v)} = \frac{\bm{\hat\varPi}(v)}{\hat\Psi(v) }.
\label{eqn_mean_theta}
\end{equation}

In a nutshell, the essence of spectroastrometric RM can be expressed in a concise equation%
\footnote{We neglect the possible nonlinear response of BLRs for simplicity. However, it is trivial to
add the nonlinear response in intensity and spectroastrometric RM (e.g., see \citealt{Li2013}).}

\begin{equation}
\left[\begin{array}{c}
F_{\rm BLR}(v, t) \\
\bm{M}_{\rm BLR}(v, t)
\end{array}\right]  =
\left[\begin{array}{c}
\Psi(v, t) \\
\bm{\varPi}(v, t)
\end{array}\right]\otimes F_{\rm c}(t),
\label{eqn_sarm_all}
\end{equation}
where $\otimes$ represents a convolution. This equation implies that variations of the emission line $F_{\rm BLR}$ and moment $\bm{M}_{\rm BLR}$
can be regarded as blurred echoes of the variations of the continuum $F_{\rm c}$. In Table~\ref{tab_variable}, we summarize the major notations and their meanings.

There had been a number of methods developed to perform deconvolution for intensity RM, among which include the maximum entropy technique (\citealt{Horne1994}), the regularized linear
inverse method (\citealt{Krolik1995,Anderson2021}), a non-parameteric Bayesian method (\citealt{Li2016}), and the Pixon-based method (\citealt{Li2021}). In addition, a Bayesian forward
approach, dynamical modeling of BLRs, had also been proposed for RM analysis
(\citealt{Pancoast2011, Pancoast2014, Li2013, Li2018}). This approach starts with a flexible dynamical model of BLRs, from which
the intensity transfer function can be directly determined, and then employs a Bayesian framework to constrain the
model parameters and hence BLR geometry and kinematics. These methods can also be applied to spectroastrometric RM.

\begin{deluxetable}{ll}
\tablecolumns{2}
\tabletypesize{\footnotesize}
\tabcaption{\centering The Description of Major Notations. \label{tab_variable}}
\tablehead{
\colhead{Notation} & \colhead{Description}
}
\startdata
$\Psi(v, \tau)$            & Intensity transfer function\\
$\tilde\Psi(\tau)$         & Velocity integral of the intensity transfer function\\
$\hat\Psi(v)$              & Delay integral of the intensity transfer function\\
$\bm{\varPi}(v, \tau)$     & Spectroastrometric transfer function\\
$\bm{\tilde\varPi}(\tau)$  & Velocity integral of the spectroastrometric transfer function\\
$\bm{\hat\varPi}(v)$       & Delay integral of the spectroastrometric transfer function\\
$F_c(t)$                   & Driving continuum flux density \\
$F'_c(t)$                  & Continuum flux density underlying the emission line\\
$F_{\rm BLR}(v, t)$             & Emission line flux density\\
$\tilde F_{\rm BLR}(t)$         & Emission line flux\\
$\bar F_{\rm BLR}(v)$           & Time averaging of emission line flux density\\
$\bm{M}(v, t)$             & Moment\\
$\bm{\tilde M}(t)$         & Velocity integral of moment\\
$\bm{\bar M}(v)$           & Time averaging of moment\\
$\bm{\varTheta}(v, t)$             & Photocenter\\
$\bm{\tilde \varTheta}(t)$         & Velocity integral of the photocenter\\
$\bm{\bar \varTheta}(v)$           & Time averaging of the photocenter\\
$\Delta\bm{\varTheta}(v, t)$            & Differential photocenter\\
$\Delta\bm{\tilde \varTheta}(t)$        & Velocity integral of the differential photocenter\\
$\Delta\bm{\bar \varTheta}(v)$          & Time averaging of the differential photocenter\\
\enddata
\end{deluxetable}

\subsection{Observational Perspectives}\label{sec_obs_per}
In practice, we can observe the SA of BLRs using a spectrometer (e.g., \citealt{Stern2015, Bosco2021}) or
interferometer (e.g., \citealt{Gravity2018}; see Section~\ref{sec_accuracy} below for a brief discussion of the observational
challenges of SA). A spectrometer yields the angular photocenters of BLRs with velocity/wavelength
along a specific direction,
related to physical photocenters through the cosmic distance of the object, namely, 
\begin{equation}
\theta(v, t) = \frac{\bm{j}\cdot\bm{\varTheta}(v, t)}{D_{\rm A}},
\label{eqn_theta}
\end{equation}
where $\theta(v, t)$ is the observed angular photocenters, $\bm{j}$ is the spatial direction of the spectrometer's slit,
and $D_{\rm A}$ is the angular-size distance.
An interferometer measures phases of BLRs with velocity/wavelength, 
related to physical photocenters through the baselines and the cosmic distance of the object, namely,
\begin{equation}
\phi(v, t) = -2\pi \frac{\bm{B}}{\lambda}\cdot\frac{\bm{\varTheta}(v, t)}{D_{\rm A}},
\label{eqn_phi}
\end{equation}
where $\lambda$ is the wavelength, $\phi(v, t)$ is the observed phases, and
$\bm{B}$ is the baseline of the interferometer. 

In realistic observations, the lights admitted to a telescope always consist of two sources: one from the continuum and the 
other from the BLR. As a result, the continuum emission also contributes to the observed photocenters, which are now 
written as 
\begin{equation}
\bm{\varTheta}(v, t) = \frac{F'_{\rm c}(t)\bm{\varTheta}_{\rm c}(v, t) + F_{\rm BLR}(v, t) \bm{\varTheta}_{\rm BLR}(v, t)}{F'_{\rm c}(t) + F_{\rm BLR}(v, t)},
\end{equation}
where $F'_{\rm c}(t)$ is the flux and $\bm{\varTheta}_{\rm c}(v, t)$ is the photocenter of the continuum underlying the emission line.
Note that usually $F'_{\rm c}(t)$ is not the same as the driving continuum $F_{\rm c}(t)$ in Equation (\ref{eqn_sarm_all}). However, AGN continuum across UV/optical and infrared bands are well correlated (e.g., \citealt{Edelson2019, Minezaki2019} and references therein). Therefore, $F'_{\rm c}(t)$ can be regarded as an echo of
$F_{\rm c}(t)$ with a time delay and possibly time blurring (if there is one). Unless stated otherwise, below we neglect this time delay and time blurring and directly use $F_{\rm c}(t)$ to replace $F'_{\rm c}(t)$ for the sake of simplicity.

It is generally reasonable to assume that the continuum's photocenter does not change with wavelength/velocity, i.e.,
$\bm{\varTheta}_{\rm c}(v, t) = \bm{\varTheta}_{\rm c}(t)$.
Therefore, we define differential photocenters to simplify the analysis
\begin{eqnarray}
\Delta \bm{\varTheta}(v, t) &=& \bm{\varTheta}(v, t) - \bm{\varTheta}_{\rm c}(t) \nonumber \\
&=& \frac{F_{\rm BLR}(v, t)}{F'_{\rm c}(t) + F_{\rm BLR}(v, t)}
\Delta\bm{\varTheta}_{\rm BLR}(v, t),
\label{eqn_dphoton}
\end{eqnarray}
where 
\begin{equation}
\Delta\bm{\varTheta}_{\rm BLR}(v, t) = \bm{\varTheta}_{\rm BLR}(v, t) -  \bm{\varTheta}_{\rm c}(t).
\end{equation}
Here, the continuum's photocenter $\bm{\varTheta}_{\rm c}(t)$ can be measured in wavelength regions without the presence of emission lines.
Below, by default, we neglect $\bm{\varTheta}_{\rm c}(t)$ and simply adopt $\bm{\varTheta}_{\rm c}(t)=0$.

It is worth pointing out that for SA observed with a spectrometer, one can alternatively first fit and subtract the continuum underlying
the emission line in the recorded 2-dimensional spectrum so as to directly measure pure photocenters of the emission line. The procedure of
continuum subtraction effectively adds extra noises to the measured pure photocenters (e.g., see \citealt{Whelan2008}).
Throughout the paper, we by default use the differential photocenters defined by Equation~(\ref{eqn_dphoton}), in which there is a scaling factor of
the flux ratio $F_{\rm BLR}/(F'_{\rm c}+F_{\rm BLR})$. Below we also simply use ``photocenter'' to refer to ``differential photocenter''.

\begin{figure*}[t!]
\centering
\includegraphics[width=0.7\textwidth]{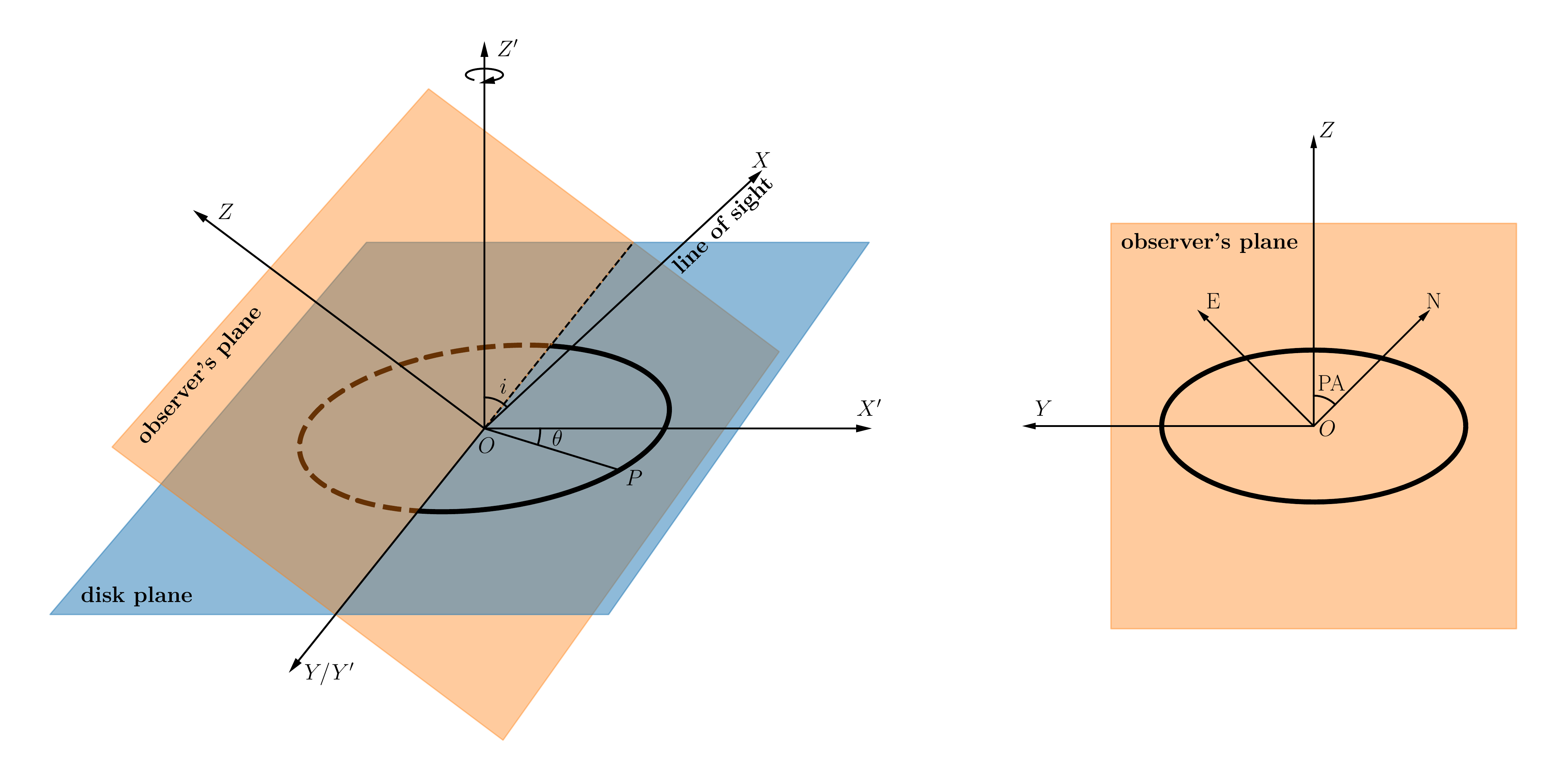}
\caption{Schematics of coordinate frames and a planar ring (left) in three-dimensional view and (right) projected in the observer's sky.
``N'' and ``E'' refer to the north and east directions, respectively, and P.A. refers to the position angle.}
\label{fig_sch}
\end{figure*}

\subsection{Cross-correlation Functions}
It is known that the cross-correlation function (CCF) between the continuum and emission line is related to the intensity transfer function as
(e.g., \citealt{Welsh1999, Li2013})
\begin{equation}
{\rm CCF}(F_{\rm c}, F_{\rm BLR}, v, \tau) =\int \Psi(v, \tau') {\rm ACF}_{\rm c}(\tau-\tau')d\tau',
\end{equation}
where the definition of CCF is given in Appendix~\ref{app_ccf} and ${\rm ACF}_{\rm c}$ represents the auto-correlation function of the continuum itself, namely,
\begin{equation}
{\rm ACF}_{\rm c}(\tau) = {\rm CCF}(F_{\rm c}, F_{\rm c}, \tau).
\end{equation}
Similarly, the CCF between the continuum and moments is related to the spectroastrometric transfer function as
\begin{equation}
{\rm CCF}(F_{\rm c}, \bm{M}_{\rm BLR}, v, \tau) =\int \bm{\varPi}(v, \tau') {\rm ACF}_{\rm c}(\tau-\tau')d\tau'.
\end{equation}

The CCF between the continuum and photocenters is not straightforward because photocenters are not linearly dependent on continuum. Nevertheless,
we can implement the following approximations. For small continuum variations ($\delta F_{\rm c}/F_{\rm c} \ll 1$), we have
\begin{eqnarray}
\bm{\varTheta}(v, t) &=& \bm{\bar \varTheta}(v) + \delta \bm{\varTheta}(v, t)\nonumber\\
&\approx& \frac{\bm{\bar M}_{\rm BLR}(v)}{\bar F_{\rm BLR}(v)}\left( 1 - \frac{F_{\rm BLR}}{\bar F_{\rm BLR}} \right)
+\frac{\bm{M}_{\rm BLR}(v,t)}{\bar F_{\rm BLR}(v)}.
\end{eqnarray}
As a result, the CCF between the continuum and photocenters can be written
\begin{eqnarray}
&&{\rm CCF}(F_{\rm c}, \bm{\varTheta}, v, \tau) \nonumber\\
&&\qquad\propto \frac{\bar F_{\rm BLR}(v)}{\sigma(F_{\rm BLR})}{\rm CCF}(F_{\rm c}, \bm{M}_{\rm BLR}, v, \tau)\nonumber \\
&&\qquad-\frac{\bm{\bar M}_{\rm BLR}(v)}{\sigma(\bm{M}_{\rm BLR})}{\rm CCF}(F_{\rm c}, F_{\rm BLR}, v, \tau),
\label{eqn_ccf}
\end{eqnarray}
where $\sigma(x)$ represents the standard deviation of the time series $x$. We can deduce from the above equation that the CCF of the photocenters is proportional to the difference between the CCFs of the moment and line profile with weights.

\begin{figure*}[t!]
\centering
\includegraphics[width=0.8\textwidth]{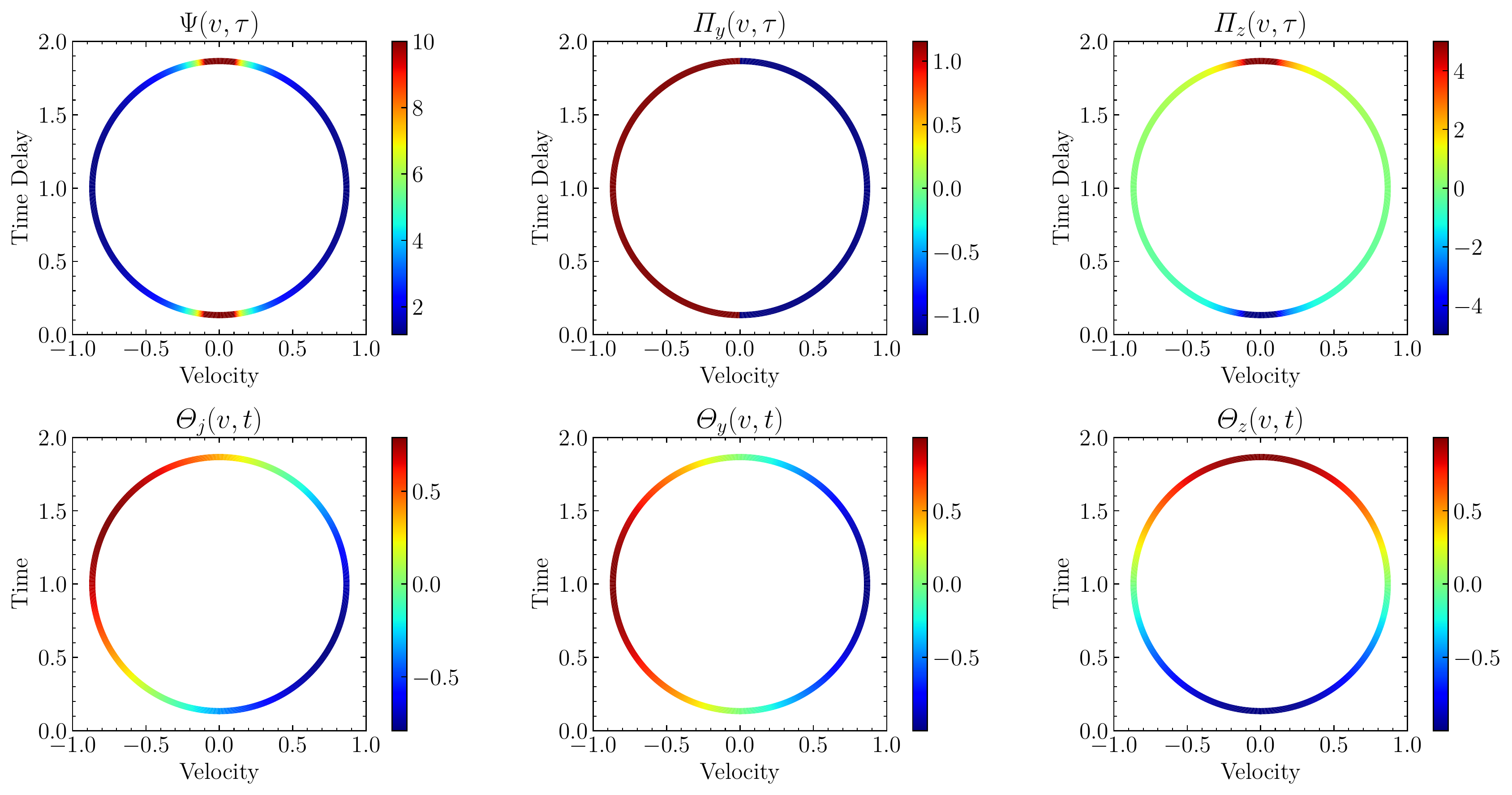}
\caption{The top three panels show the intensity transfer function $\Psi(v, \tau)$, the $y$- and $z$-components of the spectroastrometric
transfer functions $\varPi_y(v, \tau)$ and $\varPi_z(v, \tau)$ for a planar ring viewed at an inclination angle of $60^\circ$.
The bottom leftmost panel shows the photocenter $\varTheta_j(v, t)$ along the direction $j=(0, \cos45^\circ, \sin45^\circ)$.
Bottom right two panels show the photocenters $\varTheta_y(v, t)$ and $\varTheta_z(v, t)$
in a case where the continuum pulses at $t=0$. The velocity is in units of $V$ and
the time/time delay is in units of $R/c$, where $V$ is the Keplerian rotation velocity and $R$ is the radius of the ring. All color bars are in arbitrary units, but the relative scaling of the units is the same for the two spectroastrometric
transfer functions and similarly for all the photocenters.}
\label{fig_tf2d}
\end{figure*}

\begin{figure*}[!th]
\centering
\includegraphics[width=0.4\textwidth]{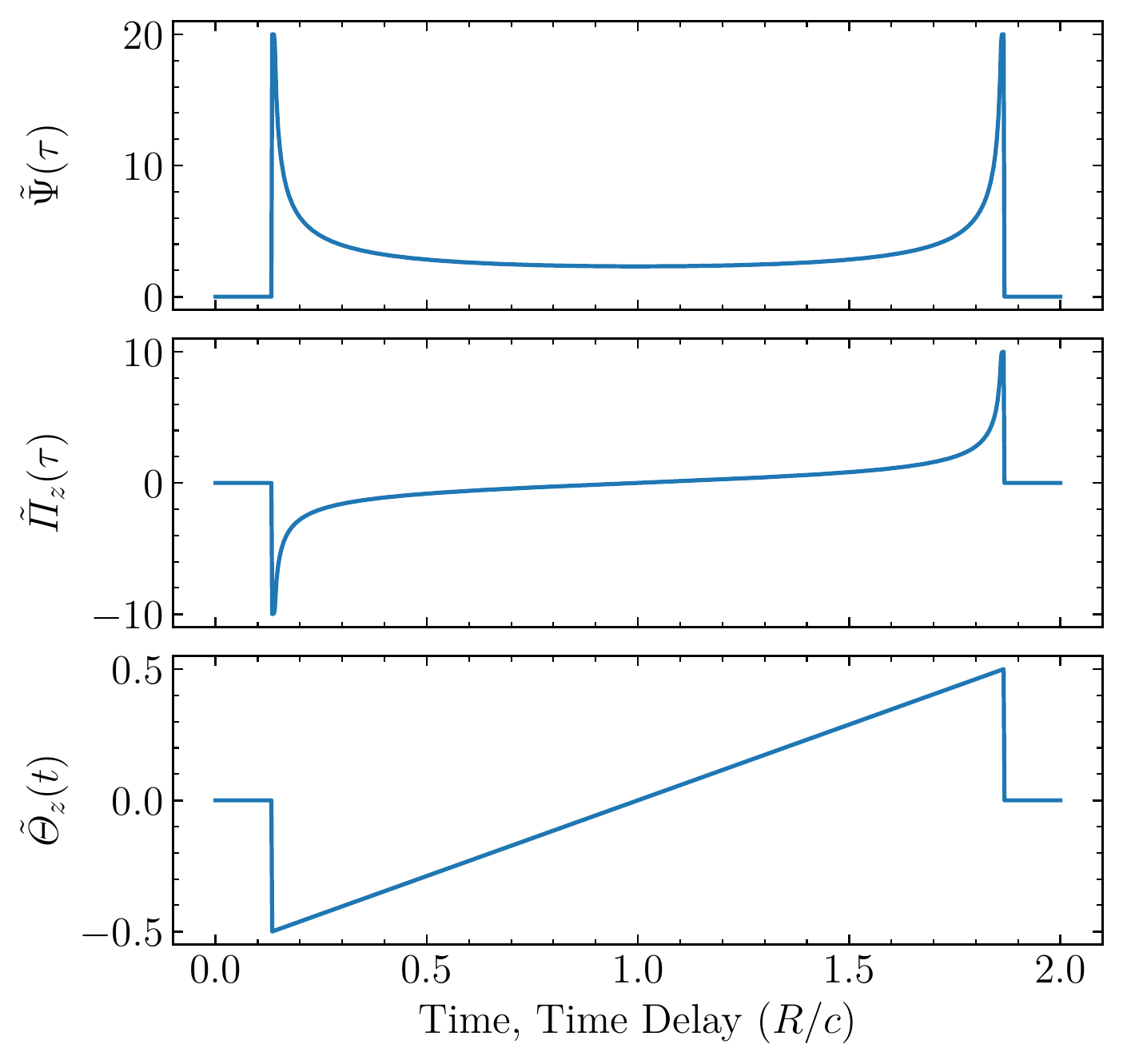}~~~~~~~~
\includegraphics[width=0.4\textwidth]{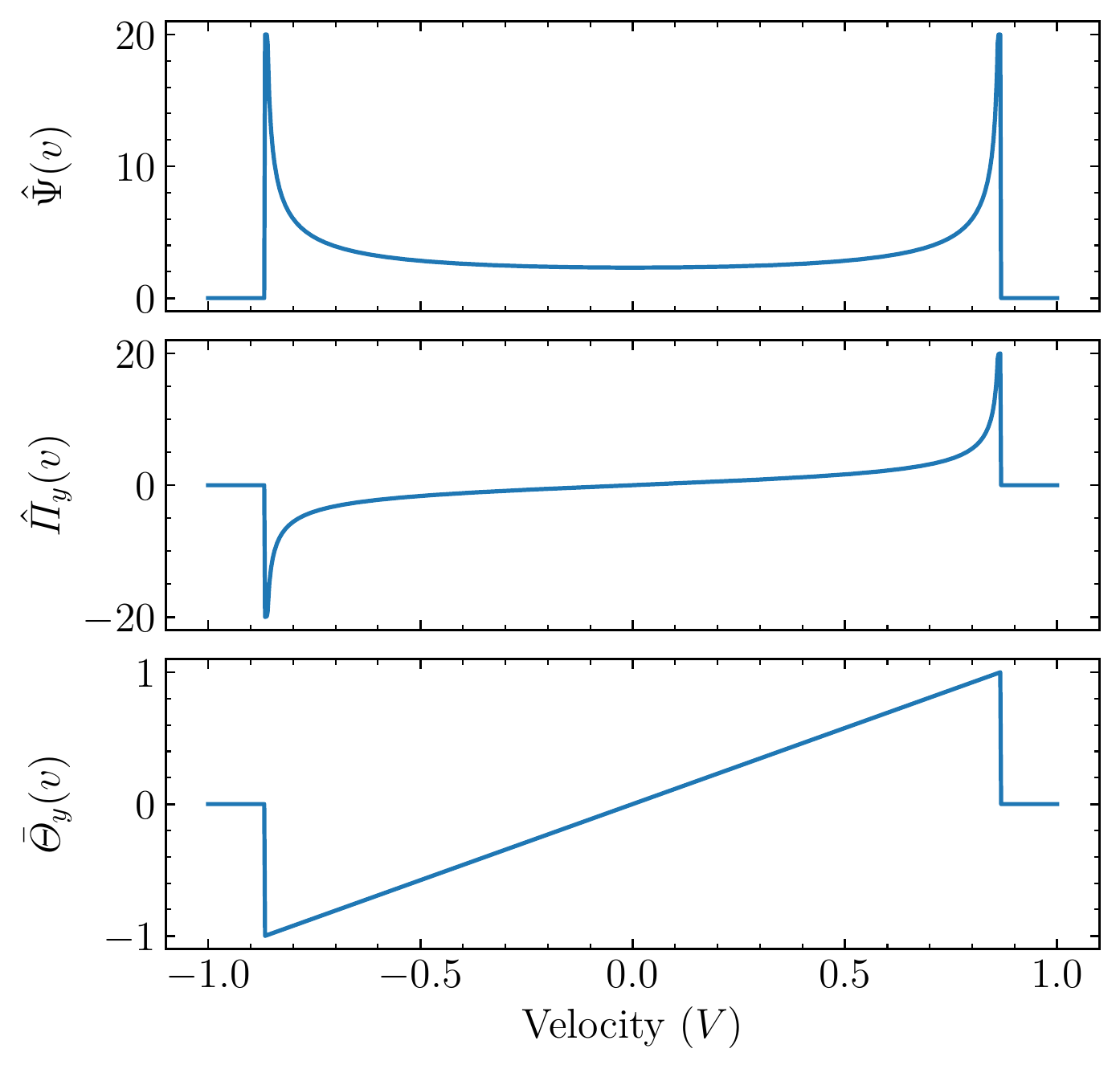}
\caption{(Left) from top to bottom are the velocity integral of the intensity transfer function $\tilde\Psi(\tau)$, and spectroastrometric
transfer function $\tilde\varPi_z(\tau)$, and the photocenter $\tilde\varTheta_z(t)$ for a planar ring as a function of time/delay.
(Right) from top to bottom are the velocity integral of the the
intensity transfer function $\hat\Psi(v)$, and spectroastrometric transfer function $\hat\varPi_y(v)$, and the photocenter $\bar\varTheta_y(v)$
as a function of velocity.
The planar ring has a radius of $R$, a Keplerian rotating velocity of $V$, and  an inclination of $60^\circ$.
The velocity is in units of $V$ and the time/time delay is in units of $R/c$. All the transfer functions are in arbitrary units.
The photocenters $\tilde\varTheta_z(t)$ and $\bar\varTheta_y(v)$ are in units of $R$.}
\label{fig_ring}
\end{figure*}

\section{Illustrative Examples}\label{sec_example}
\subsection{Two Extreme Cases}
Equations~(\ref{eqn_rm}) and (\ref{eqn_sarm}) specify how the spectral flux and photocenter of the BLR reverberate
to the continuum variations. Under a generic framework, it is not straightforward to calculate the integrals analytically in
Equations~(\ref{eqn_rm}) and (\ref{eqn_sarm}).
However, in the following two cases, there exist simple expressions.
\paragraph{Single-pulse Continuum}If the continuum only consists of a single pulses at a time $t_0$ and vanishes at other times, i.e., $F_{\rm c}(t) \propto \delta(t-t_0)$,  from Equations~(\ref{eqn_rm}) and (\ref{eqn_sarm}) we
have 
\begin{eqnarray}
	F_{\rm BLR}(v, t) &\propto& \Psi(v, t-t_0),\nonumber\\
	\bm{M}_{\rm BLR}(v, t) &\propto& \bm{\varPi}(v, t-t_0),\\
	\bm{\varTheta}_{\rm BLR}(v, t) &\propto& \frac{\bm{\varPi}(v, t-t_0)}{\Psi(v, t-t_0)},\nonumber
\end{eqnarray}
and 
\begin{eqnarray}
	F_{\rm BLR}(t) &\propto& \tilde\Psi(t-t_0),\nonumber\\
	\bm{M}_{\rm BLR}(t) &\propto& \bm{\tilde\varPi}(t-t_0),\\
	\bm{\varTheta}_{\rm BLR}(t) &\propto& \frac{\bm{\tilde\varPi}(t-t_0)}{\tilde\Psi(t-t_0)}.\nonumber
\end{eqnarray}
This implies that the variations of line flux and SA directly reflect the corresponding transfer functions.

\paragraph{Constant Continuum}Conversely, if the continuum remains constant, i.e., $F_{\rm c}(t) = F_{\rm c,0}$, the BLR's spectral flux, moment of photons,
and photocenter all do not vary with time, and
\begin{eqnarray}
	F_{\rm BLR}(v) &=& \hat\Psi(v)F_{\rm c, 0},\nonumber\\
	\bm{M}_{\rm BLR}(v) &=& \bm{\hat\varPi}(v)F_{\rm c, 0},\\
	\bm{\varTheta}_{\rm BLR}(v) &=& \frac{\bm{\hat\varPi}(v)}{\hat\Psi(v)}.\nonumber
\label{eqn_const}
\end{eqnarray}
These equations are analogous in form to Equations~(\ref{eqn_mean_fline}), (\ref{eqn_mean_mom}), and (\ref{eqn_mean_theta}) but have
different applications. Equation~(\ref{eqn_const}) refers to the cases where the continuum variations are negligible, whereas
Equations~(\ref{eqn_mean_fline}), (\ref{eqn_mean_mom}), and (\ref{eqn_mean_theta}) refer to time averaging so that any time-dependent information
is eliminated. When there is only one-epoch observation, one can use the above equations as the first order of approximation.

\begin{deluxetable*}{ccccl}
\tablecolumns{5}
\tabletypesize{\footnotesize}
\tabcaption{\centering BLR Model Parameters and Priors Used for Simulation Data. \label{tab_param}}
\tablehead{
\colhead{Parameter} & \colhead{Value} & \colhead{Unit} & \colhead{Prior} & \colhead{Description}
}
\startdata
$M_\bullet$        & $2.6\times10^{8}$ & $M_\odot$        & LogUniform(10$^7$, 10$^{10}$)      & Black hole mass \\
$R_{\rm BLR}$      & 146               & light-day        & LogUniform(10, $\Delta T$/2)  & Mean BLR radius \\
$\beta$            & 1.4               & \nodata          & Uniform(0, 2)                       & Shape parameter of the radial distribution of BLR clouds  \\
$F$                & 0.24              & \nodata          & Uniform(0, 1)                       & Inner edge of the BLR in units of $R_{\rm BLR}$\\
$\theta_{\rm opn}$ & 45                & Degree           & Uniform(0, 90)                      & Opening angle of the BLR\\
$\theta_{\rm inc}$ & 12                & Degree           & Uniform($\cos0^\circ$, $\cos90^\circ$) & Inclination angle of the BLR\\
$D_{\rm A}$        & 550               & Mpc              & LogUniform(10, 10$^4$) & Angular-size distance\\
PA                 & 0                 & Degree (E of N)  & Uniform($-180^\circ$, $180^\circ$) & Position angle on the sky
\enddata
\tablecomments{$\Delta T$ represents the time length of the continuum light curve.
``Uniform'' refers to a uniform prior and ``LogUniform'' refers to a uniform prior for the logarithm of the parameter. }
\end{deluxetable*}
\begin{figure*}[t!]
\centering
\includegraphics[width=0.8\textwidth]{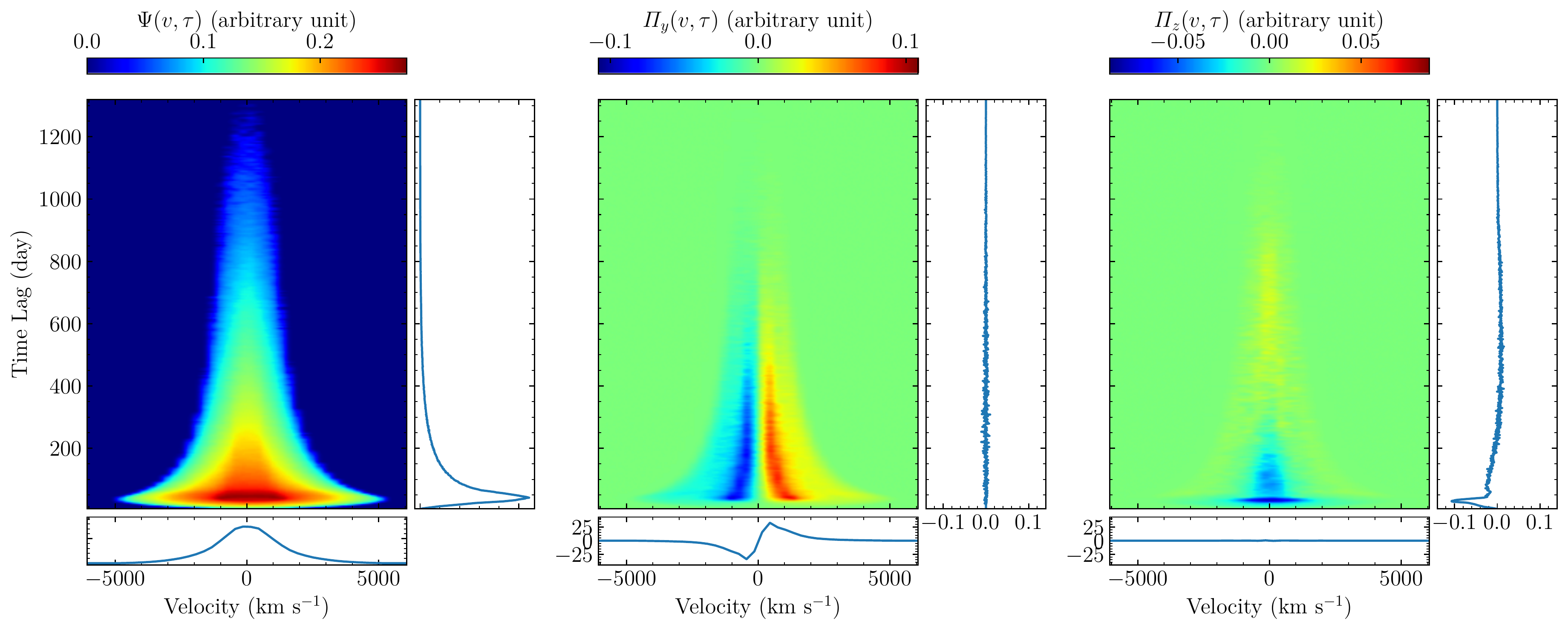}
\caption{From left to right are the intensity transfer function $\Psi(v, \tau)$ and spectroastrometric transfer functions $\varPi_y(v, \tau)$ and $\varPi_z(v, \tau)$
for a general BLR model with the model parameters specified in Table~\ref{tab_param}. In each panel, the right and bottom subpanels show the velocity
and delay integrals of the transfer functions, respectively. All these transfer functions are in arbitrary units, but the relative scaling of the units  is the same for
$\varPi_y(v, \tau)$ and $\varPi_z(v, \tau)$.}
\label{fig_sim_tf}
\end{figure*}

\begin{figure*}
\centering
\includegraphics[width=0.8\textwidth]{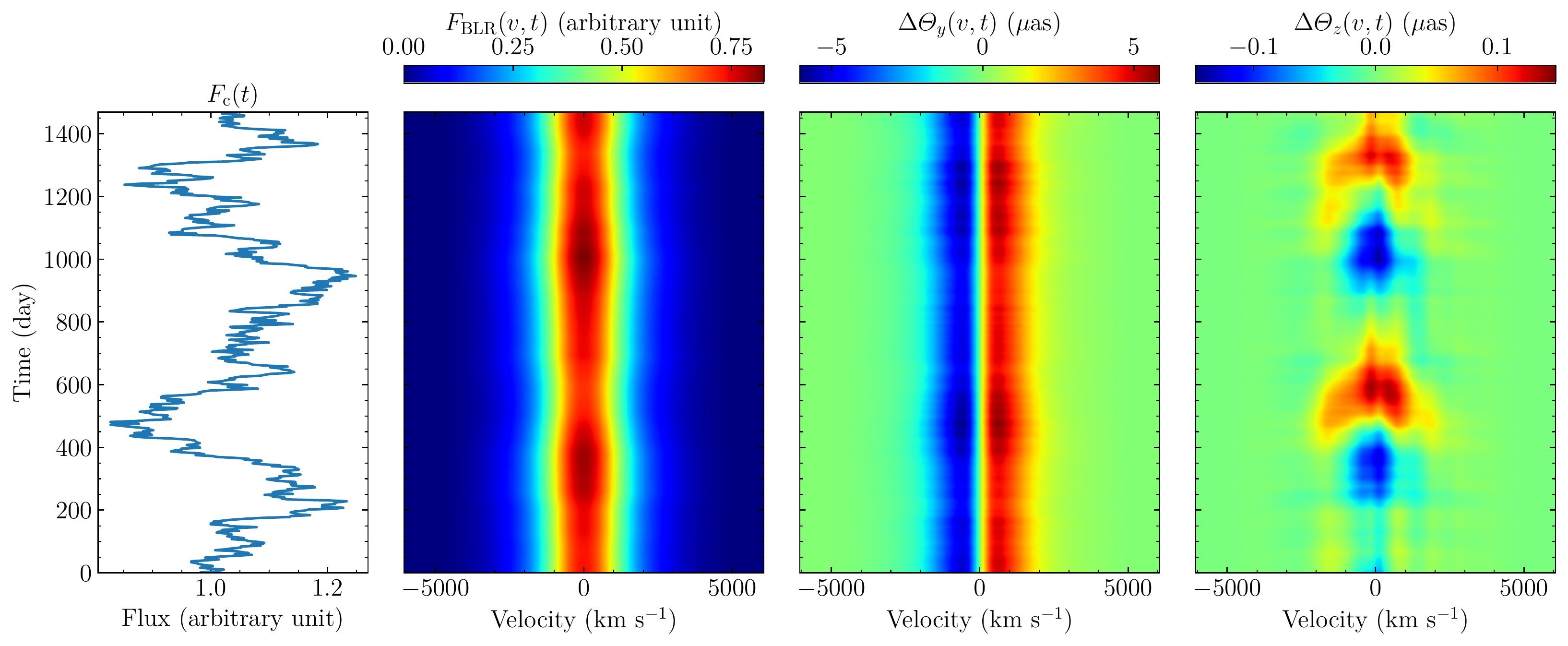}
\caption{From left to right are a simulated continuum light curve $F_{\rm c}(t)$, time series of the emission line $F_{\rm BLR}(v, t)$, and photocenters
$\Delta\varTheta_y(v,t)$ and $\Delta\varTheta_z(v, t)$.}
\label{fig_sim_ts}
\end{figure*}

\subsection{An Inclined Planar Ring}
For an inclined planar ring, there exist analytical expressions for the intensity and spectroastrometric transfer functions.
We create a Cartesian coordinate frame $X'Y'Z'$ with its origin located at the center of the ring
and $Z'$-axis aligned with its rotating axis. We then rotate the $X'Y'Z'$ frame around the $Y'$-axis
by an angle of $\pi-i$ to create a new Cartesian coordinate frame $XYZ$. Here, $i$ is the inclination angle of the
ring. We set the LOS along the $X$-axis so that the $YZ$ plane defines the observer's sky plane.
For simplicity, we assume that the emissivity $\epsilon$ is isotropic and constant along the ring.
Figure~\ref{fig_sch} shows a schematic of the coordinates and the planar ring. In Appendix~\ref{app_ring}, we derive the analytical expressions of
spectroastrometric RM for an inclined planar ring.

In the three top panels of Figure~\ref{fig_tf2d}, we plot the intensity transfer function $\Psi(v, \tau)$, the $y$- and $z$-components of the
spectroastrometric transfer function $\varPi_y(v, \tau)$ and $\varPi_z(v, \tau)$ for a planar ring with an inclination angle of $60^\circ$.
The two bottom right panels of Figure~\ref{fig_tf2d} show the photocenters $\varTheta_y(v, t)$ and $\varTheta_z(v, t)$ in a case where
the continuum pulses at $t=0$. The bottom leftmost panel shows the photocenter along the direction $\bm{j}=(0, \cos\pi/4, \sin\pi/4)$.
All these quantities are nonzero only along the ellipse $v^2/V^2 + (c\tau/R-1)^2=\sin^2i$. It is easy to show that for a BLR composed
of a series of coplanar rings, the above plots are just a superimposition of corresponding ellipses with different velocities $V$ and radius $R$.
In Figure~\ref{fig_ring}, we illustrate velocity integrals and delay integrals of the transfer functions.
In the left bottom panel, we plot the change in the photocenter $\tilde\varTheta_z(t)$ with time in a case where the continuum pulses at $t=0$.
In the right bottom panel, we plot the photocenter $\hat\varTheta_y(v)$ with velocity in a case where the continuum is a constant.

\begin{figure}[t!]
\centering
\includegraphics[width=0.48\textwidth]{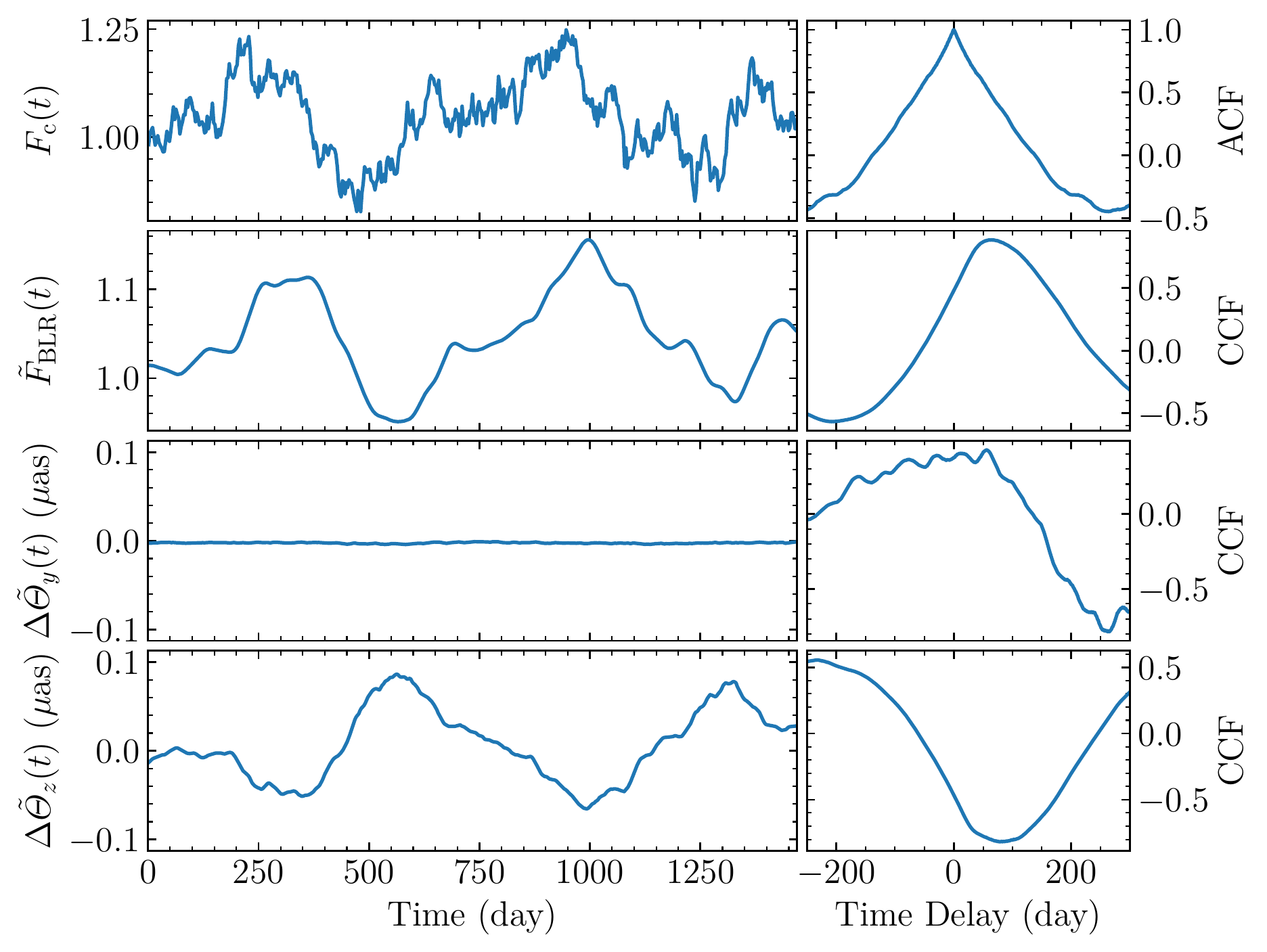}
\caption{(Left) From top to bottom are simulated light curves of the continuum, emission line, and $y$- and $z$-component velocity integrals of the photocenters.
The units of $F_{\rm c}$ and $\tilde F_{\rm BLR}$ are arbitrary, but the relative scaling is the same. (Right) From top to bottom are the ACF of the continuum light curve and the CCFs between the continuum and the corresponding light curves plotted in the left panels.}
\label{fig_sim_lc}
\end{figure}

\begin{figure*}[t!]
\centering
\includegraphics[width=0.9\textwidth]{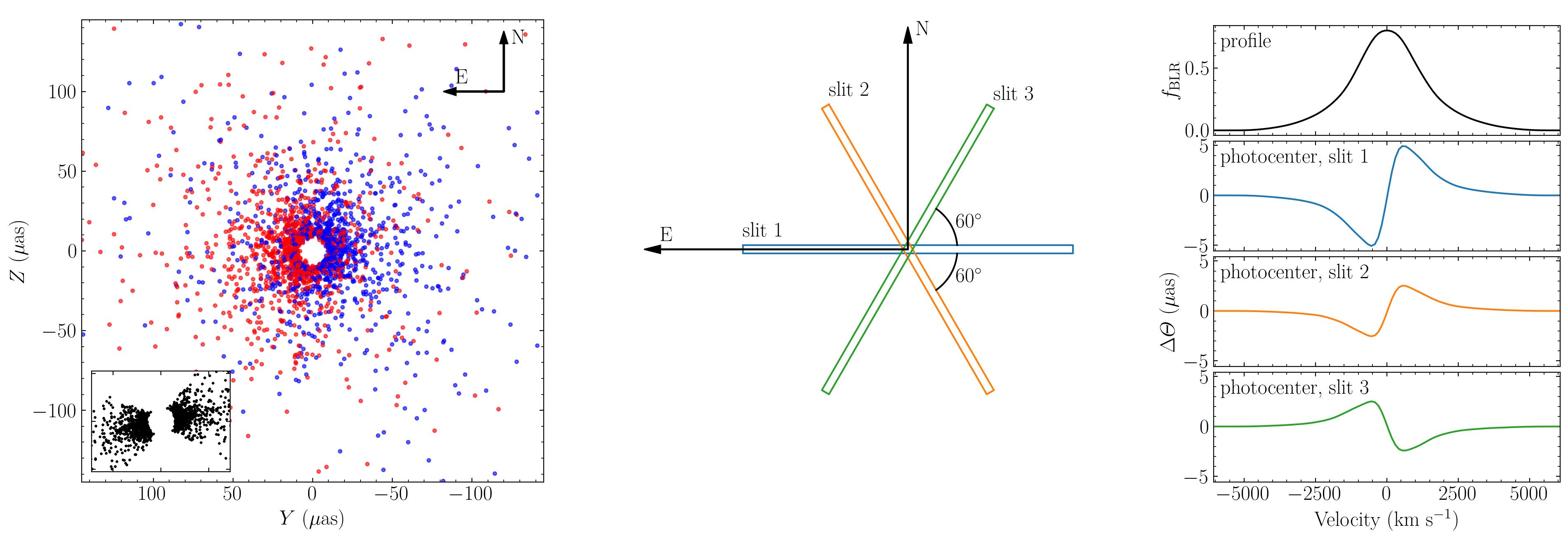}
\caption{(Left) Distribution of the BLR clouds projected to the sky, in which the inset shows the edge-on cutaway ($Z$-$X$) view of the BLR clouds. The red and blue colors mean receding and approaching clouds, respectively.
(Middle) A schematic of three slit orientations that are rotated by 60$^\circ$ from one another. ``N'' and ``E'' refer to the north
and east directions, respectively. (Right) The line profile (arbitrary unit) and photocenters at one epoch are generated using the parameter values listed in Table~\ref{tab_param}.}
\label{fig_slit}
\end{figure*}

\begin{figure*}[!tp]
\centering
\includegraphics[width=0.8\textwidth]{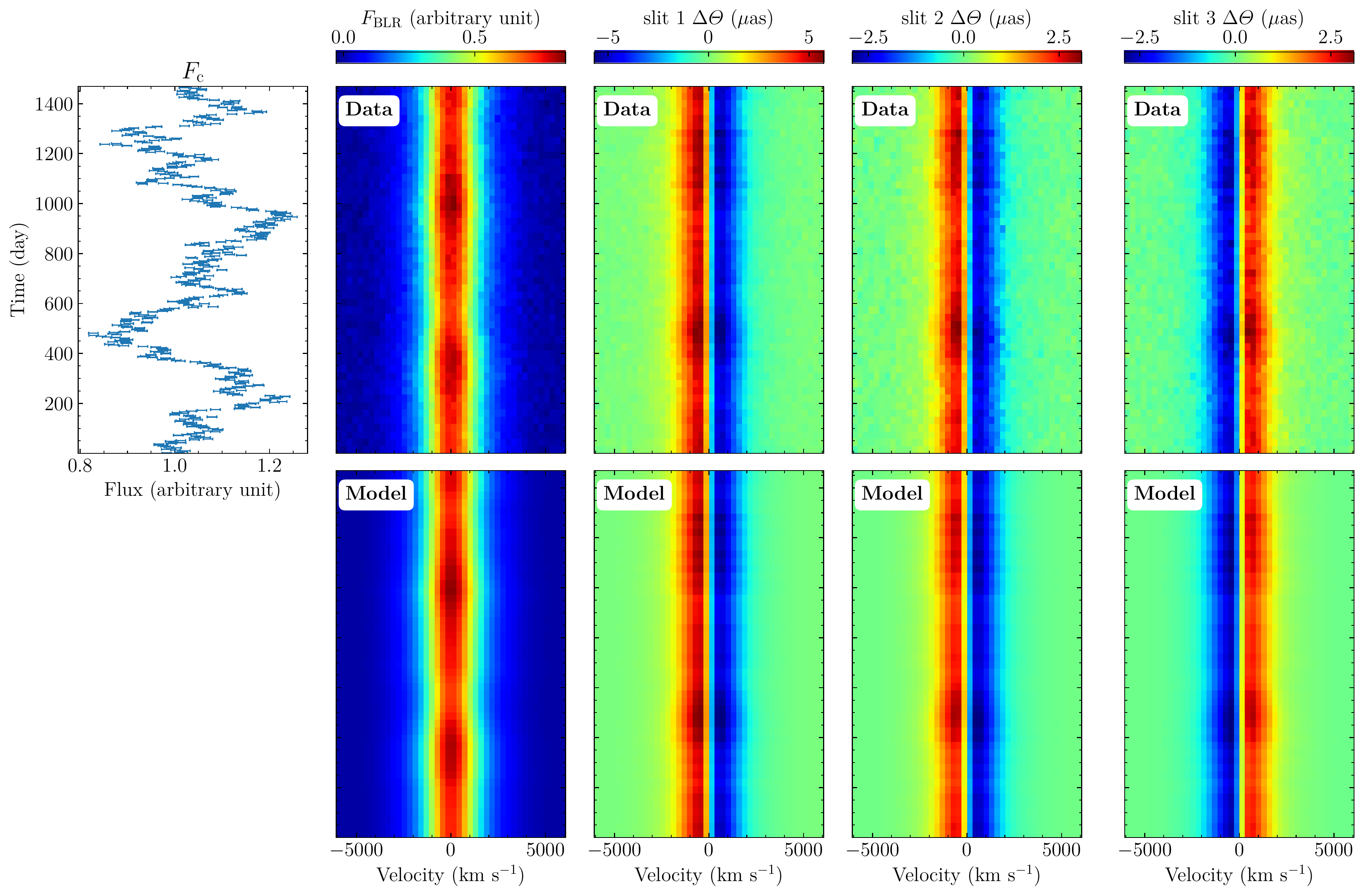}
\caption{Simulated data for the continuum ($F_{\rm c}$), line profile ($F_{\rm BLR}$), and photocenters ($\Delta\varTheta$) at three slit orientations (illustrated in Figure~\ref{fig_slit}). The bottom panels show reconstructions with the BLR model. }
\label{fig_sim_sarm_data}
\end{figure*}

\section{A General Case of BLRs}\label{sec_generic}
In this section, we show how to calculate the spectroastrometric RM signal for a general case of BLRs. To this end, we need to (1) generate mock AGN continuum
light curves and (2) construct a BLR model.
We assume that the continuum variations follow the damped random walk (DRW) model and generate mock light curves using the procedure detailed in Appendix~\ref{app_lc}.
Regarding the BLR model, without losing the generality,
we simply assume a disk-like axisymmetric geometry and Keplerian rotation. The BLR consists of a large number of discrete point-like clouds, which rotate coherently
around the central SMBH and reprocess the ionizing continuum. The clouds' emissions are isotropic and the shadowing among clouds is neglected for simplicity.
By using cylindrical coordinates, the clouds follow a gamma distribution in the radial $r$-direction, parametrized by the mean BLR radius $R_{\rm BLR}$,
the inner edge parameter $F$, and the shape parameter $\beta$ (see \citealt{Pancoast2014} and \citealt{Li2018} for details); in the vertical $\theta$ direction,
the clouds subtend an opening angle $\theta_{\rm opn}$ and have a uniform distribution in terms of $\theta$; in the azimuthal $\varphi$ direction, the clouds are also
distributed uniformly. The BLR is viewed at an inclination angle of $\theta_{\rm inc}$. In Table~\ref{tab_param}, we
summarize the BLR model parameters and list their values used in our calculations. The flux ratio of the line peak to the continuum
is set to about 0.8 (see Equation~\ref{eqn_dphoton}). The spectral broadening is set to 235 km~s$^{-1}$ to account for instrumental broadening effects. These values
are consistent with the inferences from the spectroastrometric observations of 3C~273 by the \cite{Gravity2018}. In Appendix~\ref{app_sa}, we demonstrate how to calculate the transfer functions (Equations~\ref{eqn_stf} and \ref{eqn_satf}) given a BLR model.

\begin{figure*}[t!]
\centering
\includegraphics[width=0.7\textwidth]{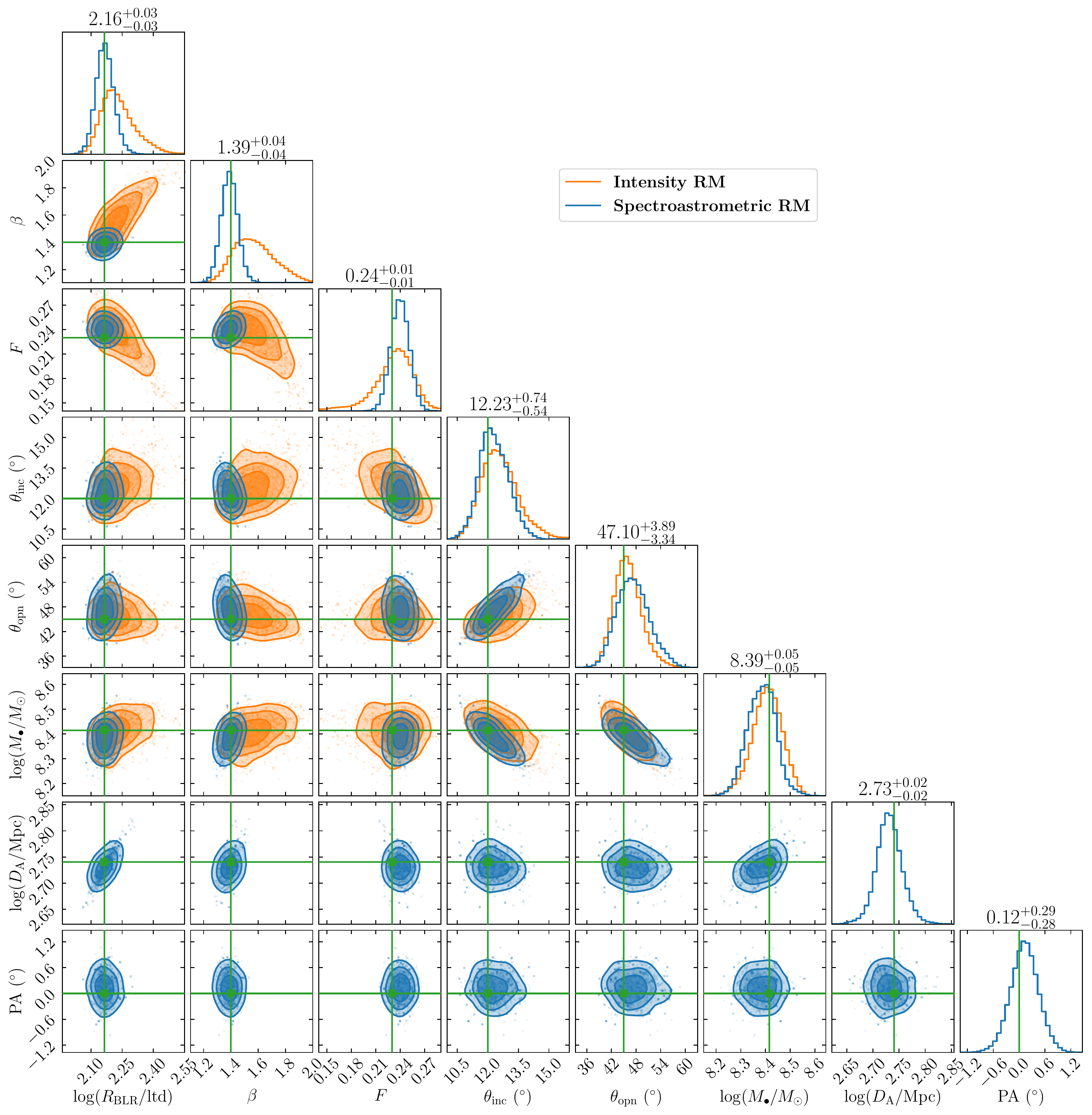}
\caption{The inferred posterior distributions of model parameters using spectroastrometric RM data (blue) and only intensity RM data (yellow). Green lines represent the input values. The contours are at the 1$\sigma$, 1.5$\sigma$, and 2$\sigma$ levels. The numbers above each diagonal panel mark the media value and 68.3\% confidence intervals from the spectroastrometric RM data. }
\label{fig_hist}
\end{figure*}

In Figure~\ref{fig_sim_tf}, we show the obtained intensity transfer function $\Psi(v, \tau)$ and spectroastrometric transfer functions $\varPi_y(v, \tau)$ and $\varPi_z(v, \tau)$, as well as their corresponding delay and velocity integrals. The intensity transfer function has a bell-like shape, with the significant
responses concentrated around 50 days and a long tail extending to several hundred days. This is because the radial distribution of the BLR clouds
has a steep decay with radius for the shape parameter $\beta=1.4$. The $y$-component of the spectroastrometric transfer function $\varPi_y$ has an
$S$-shape along the velocity axis, arising from the Keplerian rotation that causes the redshifting and blueshifting clouds to offset oppositely on the sky.
The velocity integral $\tilde\varPi_y(\tau)$ goes to zero because the red and blue parts exactly cancel out. The $z$-component $\varPi_z$ displays a different response
pattern with $\tilde\varPi_y$. The nearside of the BLR responds earlier and has a negative $z$-coordinate (see Figure~\ref{fig_sch}); therefore, $\tilde\varPi_z$ is negative
at short time delays. At long time delays, the farsize of the BLR starts responding and $\tilde\varPi_z$ turns positive. This also leads the delay integral $\hat\varPi_z(v)$
to vanish because the parts with short and long delays cancel out in the integral.

In Figure~\ref{fig_sim_ts}, we plot a randomly generated continuum light curve and its driven time series of the emission line and photocenters.
As expected, the variations in the emission line flux are delayed with respect to those of the continuum. Regarding the photocenters,
first of all, we note that while the intensity transfer function is always positive, the spectroastrometric transfer function can be either positive or negative
because, as mentioned above, the defined photocenters of BLR clouds can either be positive or negative (see Figure~\ref{fig_sch}).
This is also seen in the velocity integrals of the line flux and photocenters shown in Figure~\ref{fig_sim_lc}.
The line flux $\tilde F_{\rm BLR}(t)$ varies as a delayed and blurred echo of the continuum and a positive CCF peak appears around 50 days
(because the intensity transfer function peaks at about 50 days; see Figure~\ref{fig_sim_tf}).
The $y$-component $\Delta\tilde \varTheta_y(t)$ is almost zero since the photocenter has positive values at the blue wavelength and negative values at the red wavelength so that their velocity integral cancel out. The $z$-component $\Delta\tilde\varTheta_z(t)$ displays an inverse variation pattern compared to that of the continuum due to
the strong negative response of $\tilde \varPi_z(t)$ around 50 days (see Figure~\ref{fig_sim_tf}), which leads to a negative CCF as shown in the right bottom panel of Figure~\ref{fig_sim_lc}. However, we note that such negative CCFs are not intrinsic and just caused by the definition of the photocenter axes shown in Figure~\ref{fig_sch}.

From Figure~\ref{fig_sim_ts}, we can also find that the overall variation amplitude of the $y$-component photocenters $\Delta\varTheta_y$ around the line core is at a level of several microarcseconds, depending on the continuum variability, BLR size, and angular-size distance as well. Generally speaking, larger continuum variability, a larger BLR, or a smaller angular-size distance will yield larger variations in photocenters. However, a larger BLR usually corresponds to a more luminous AGN and thereby a longer variation timescale.

\begin{figure*}[t!]
\centering
\includegraphics[width=0.7\textwidth]{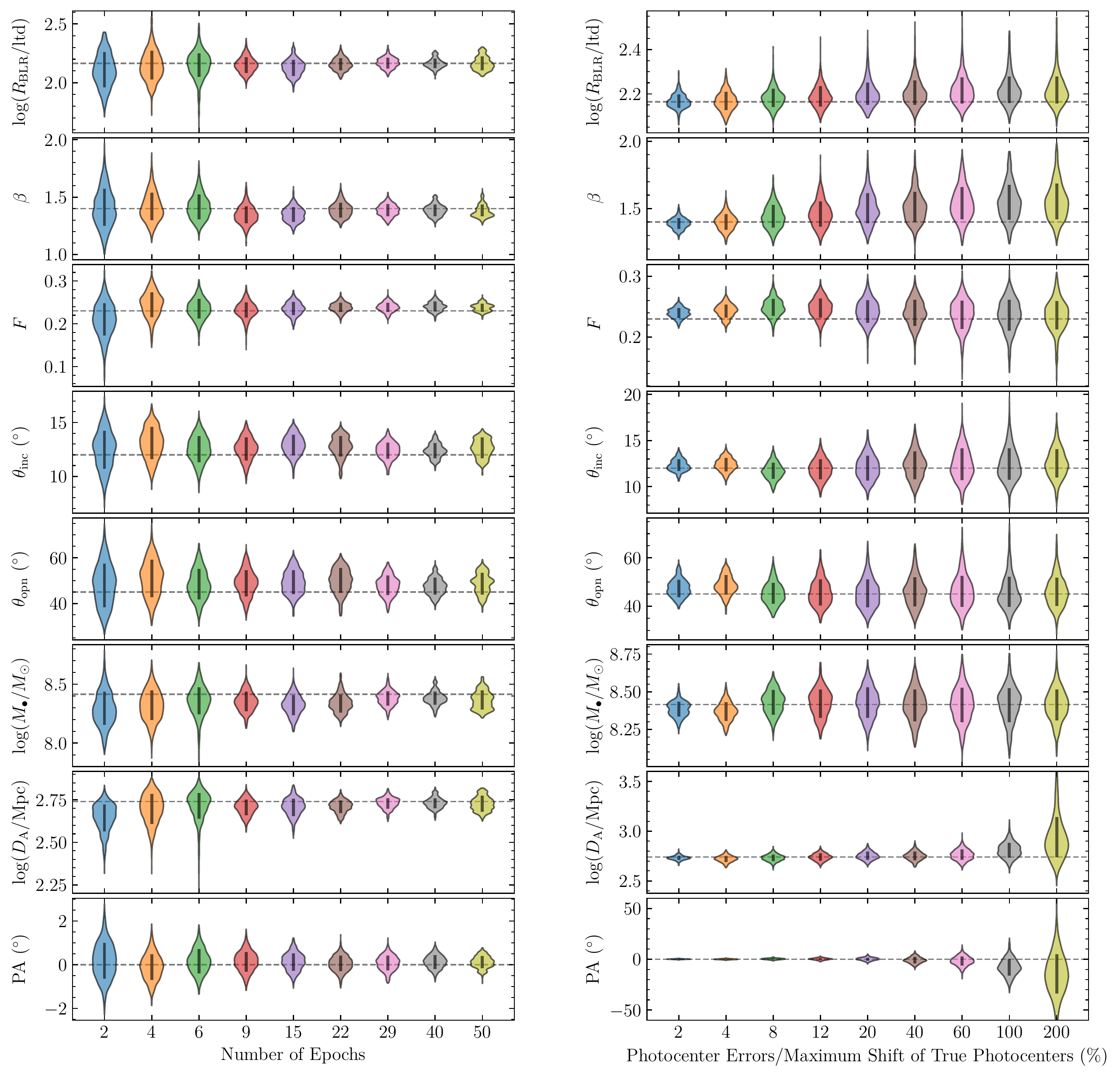}
\caption{Violin plots for the inferred parameter values and uncertainties with (left) the number of epochs and (right) the injected errors to the photocenters.
The maximum shift of the true photocenters is $\sim$5 $\mu$as.
The vertical black lines inside the violin plots represent the 68.3\% confidence intervals. Note that the sampling of continuum light curves is the same in all cases.}
\label{fig_para_values}
\end{figure*}

\section{Bayesian Inferences}\label{sec_bayes}
\subsection{A Bayesian Framework}
We now develop a generic Bayesian framework to infer BLR parameters from spectroastrometric data.
Given a DRW model and BLR dynamical model,
we can reconstruct the continuum light curve from the observed continuum light curve and calculate the spectroastrometric signals
using the procedures described in the preceding sections. The calculated spectroastrometric signals are
then compared against the observed data, namely, the flux ($\bm{D}_{\rm line}$) and SA ($\bm{D}_{\rm sa}$)
data of the emission line. By assuming that the data errors are Gaussian and uncorrelated, the likelihood probability is given by
\begin{equation}
P(\bm{D}_{\rm line}, \bm{D}_{\rm sa}|\bm{\theta}) = P(\bm{D}_{\rm line}|\bm{\theta})P(\bm{D}_{\rm sa}|\bm{\theta}),
\end{equation}
where $\bm{\theta}$ represents the model parameter set,
\begin{equation}
P(\bm{D}_{\rm line}|\bm{\theta}) = \prod_{ij} \frac{1}{\sqrt{2\pi}\sigma_{{\rm line}, ij}}
\exp\left\{-\frac{[F_{ij}-F^{m}_{ij}(\bm{\theta})]^2}{2\sigma^2_{{\rm line}, ij}} \right\},
\end{equation}
and
\begin{equation}
P(\bm{D}_{\rm sa}|\bm{\theta}) = \prod_{ij} \frac{1}{\sqrt{2\pi}\sigma_{{\rm sa}, ij}}
\exp\left\{-\frac{[\bm{\varTheta}_{ij}-\bm{\varTheta}^{m}_{ij}(\bm{\theta})]^2}{2\sigma^2_{{\rm sa}, ij}} \right\},
\end{equation}
where the superscript ``$m$'' represents the corresponding quantities calculated from the BLR model
and $i$ and $j$ represent the epoch and wavelength bin.
The posterior probability is then given by
\begin{equation}
P(\bm{\theta} | \bm{D}_{\rm line}, \bm{D}_{\rm sa}) = \frac{P(\bm{D}_{\rm line}, \bm{D}_{\rm sa}|\bm{\theta})P(\bm{\theta})}{P(\bm{D}_{\rm line}, \bm{D}_{\rm sa})},
\end{equation}
where $P(\bm{\theta})$ is the prior probability of model parameters and $P(\bm{D}_{\rm line}, \bm{D}_{\rm sa})$
is the Bayesian evidence.

We employ the Markov Chain Monte Carlo (MCMC) technique to optimize the posterior probability and the
diffusive nested sampling algorithm (\citealt{Brewer2011}) to generate Markov chains.
We implement the above procedures based on our previously developed package \texttt{BRAINS} for BLR dynamical
modeling (\citealt{Li2013, Li2018, Li2022}), which is publicly available at \url{https://github.com/LiyrAstroph/ BRAINS}.
This package is written in C language and uses the diffusive nested sampling library \texttt{CDNest} (\citealt{Li2020cdnest}) based on the original work of the diffusive nested sampling algorithm by \cite{Brewer2011}. This package supports the standardized message-passing interface and therefore can run on a wide range of supercomputer clusters without any reliance on special features of proprietary compilers.

\subsection{Validity Tests}\label{sec_tests}
\subsubsection{Simulation Configurations}
We generate simulated datasets by injecting Gaussian noise to mimic realistic observations and then run our Bayesian package
to test its validity.
As mentioned in Section~\ref{sec_obs_per}, there are two approaches to measure the SA of BLRs, namely, using
a spectrometer (e.g., \citealt{Bosco2021}) or an interferometer (e.g., \citealt{Gravity2018}).
Their respective observables, the photocenter and phase, are indeed related through Equations~(\ref{eqn_theta})
and (\ref{eqn_phi}). Hereafter, unless stated otherwise, we only use photocenters for our following simulation tests.

As illustrated in the middle panel of Figure~\ref{fig_slit}, we generate spectroastrometric data under three slit orientations rotated by 60$^\circ$ from one another at each epoch (see, e.g., \citealt{Pontoppidan2008}). Without loss of generality, we set the three orientations to $\bm{j}=(1, 0)$, $(\cos\pi/3, \sin\pi/3)$, and $(\cos2\pi/3, \sin2\pi/3)$. We neglect the possible errors in slit positioning and assume that the orientations remain stable for all epochs. We adopt the values of the model parameters listed in Table~\ref{tab_param} and show the generated distributions of BLR clouds on the sky in the left panel of Figure~\ref{fig_slit}. In the right panel of Figure~\ref{fig_slit}, we plot the line profile and photocenters as a function of wavelength at the three orientations for a randomly selected epoch.
To account for measurement errors, we add Gaussian noises with a standard deviation of 0.01 (arbitrary unit) for the line profiles and 0.1 $\mu$as for the photocenters.
As a reference, the peak flux is about 0.8 and the maximum shift of the true photocenters is about 5 $\mu$as (see Figure~\ref{fig_slit}), which correspond to a signal-to-noise ratio (S/N) of 80 and 50, respectively. We use 42 equally spaced velocity bins over a range between -6000 and 6000 km~s$^{-1}$ (corresponding to a spectral resolution of $\lambda/\Delta\lambda\approx 1000$). The spectral instrumental broadening is set to 235 km~s$^{-1}$.

The continuum light curve is generated using the DRW model with a typical timescale of 300 days and
a variation amplitude parameter that results in an overall relative variability of about 40\%. The mean continuum flux density is set to be unity, resulting in a flux ratio of $F_{\rm BLR}/F'_c\approx0.8$ at the line peak. Again, Gaussian noise with a standard deviation of 1\% is injected into the continuum light curve. The time span is set to 1500 days and the sampling cadence is set to 3 days apart. Such a cadence is feasible considering the fact that one can synthesize data from different monitoring campaigns as well as from public time-domain surveys, such as the All-Sky Automated Survey for Supernovae (\citealt{Kochanek2017}) and the Zwicky Transient Facility (\citealt{Graham2019}). The seasonal gaps are not included for the sake of simplicity.

We stress that the above configurations are designed only for illustration purposes. In particular, the injected S/N of the photocenters is somehow idealized considering the astrometric accuracy achievable at current facilities (see Section~\ref{sec_accuracy} below). Besides, there are several additional factors that are not included in simulations. (1) We use uniform errors for all epochs. In reality, the errors might vary among epochs because of different observing conditions; (2) We do not include the narrow-line component superimposed on the broad emission line, which might affect the observed photocenters, depending on its flux ratio compared to the broad component. A possible economic solution is masking out the wavelength range with the narrow line in the BLR modeling; (3) AGNs show a wide range of variability, not all of which are conducive to doing spectroastrometric RM analysis. This issue can be resolved by the preselection of targets based on variability. Detailed investigations into these factors are quite beyond the scope of this work. Below we will only show how the two key configurations, namely, sampling rate and photocenter errors, affect the Bayesian inference.

\begin{figure*}[t!]
\centering
\includegraphics[width=0.7\textwidth]{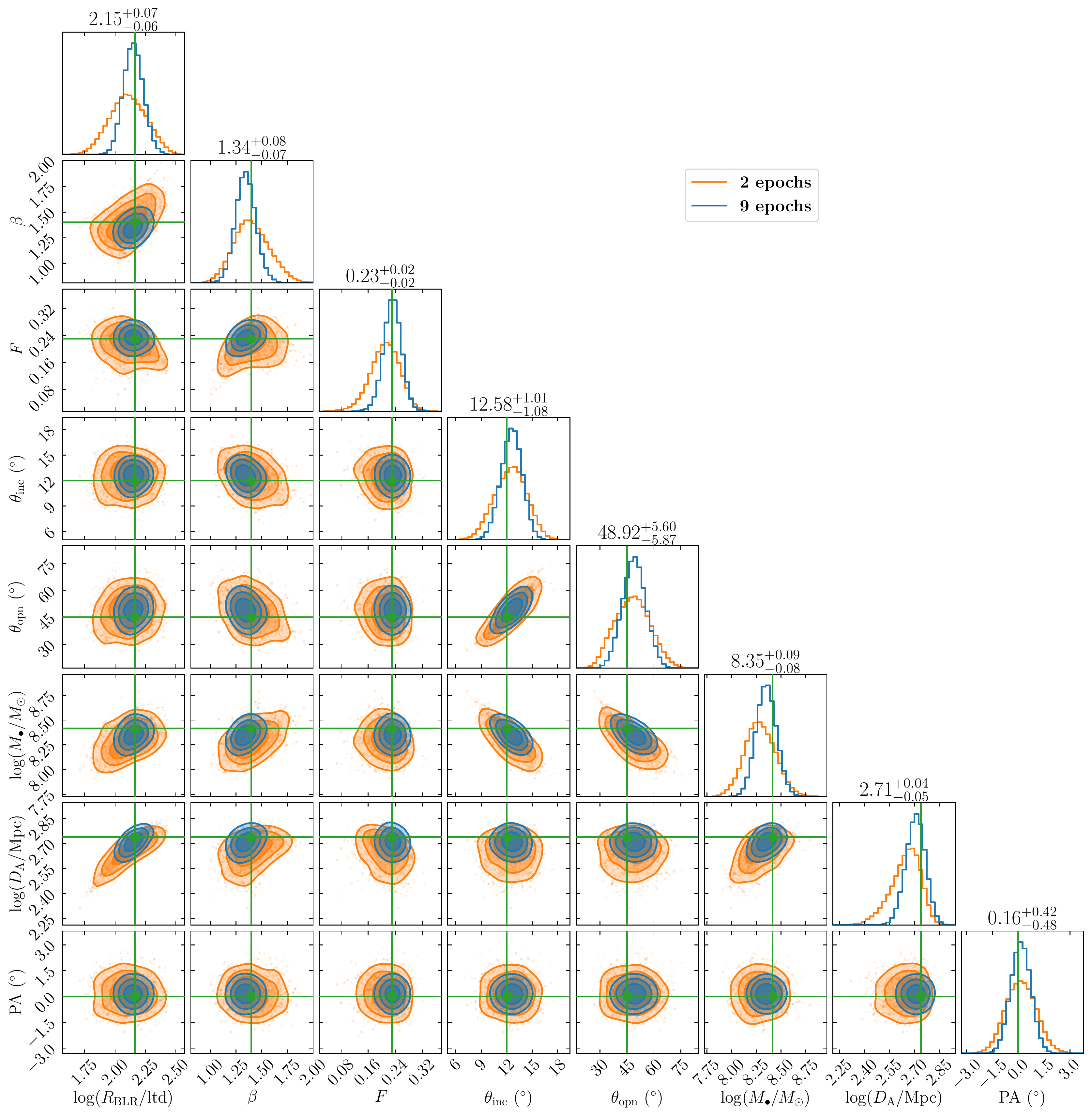}
\caption{A comparison of the inferred posterior distributions of model parameters for nine (blue) and two (yellow) randomly selected epochs  of spectroastrometric RM data.
Green lines represent the input values. The contours are at the 1$\sigma$, 1.5$\sigma$, and 2$\sigma$ levels. The numbers above each diagonal panel mark the media value and 68.3\% confidence intervals from the case of nine epochs. Note that the sampling of continuum light curves are the same for both cases of two and nine epochs.}
\label{fig_comp_epoch}
\end{figure*}

\subsubsection{A Test Case}
As a test case, we generate 50 equally spaced epochs of SA over a time span of 1500 days. Figure~\ref{fig_sim_sarm_data} shows the generated
mock data of the continuum ($F_{\rm c}$), line profile ($F_{\rm BLR}$), and photocenters ($\varTheta$) at three slit orientations (illustrated in Figure~\ref{fig_slit}).
Figure~\ref{fig_hist} plots the posterior distributions of the model parameters. As can be seen, all the parameters are well consistent with the input values at a level of 1$\sigma$ confidence. In particular, both the black hole mass and angular-size distance are well constrained. For the sake of comparison, in the bottom panels of Figure~\ref{fig_sim_sarm_data}, we plot the recovered line profile and photocenters, which are again well consistent with the simulated data.
In Figure~\ref{fig_hist}, we also superimpose the posterior distributions only using the simulated spectral data (namely, intensity RM). The obtained parameters have relatively broader contours, in particular, for the parameters $R_{\rm BLR}$ and $\beta$, reflecting the potential of spectroastrometric RM in constraining BLR geometry and kinematics. In addition, if we use a more stringent prior for the angular-size distance from the standard cosmology, all parameter inferences can be further improved.

\begin{figure}[t!]
\centering
\includegraphics[width=0.45\textwidth]{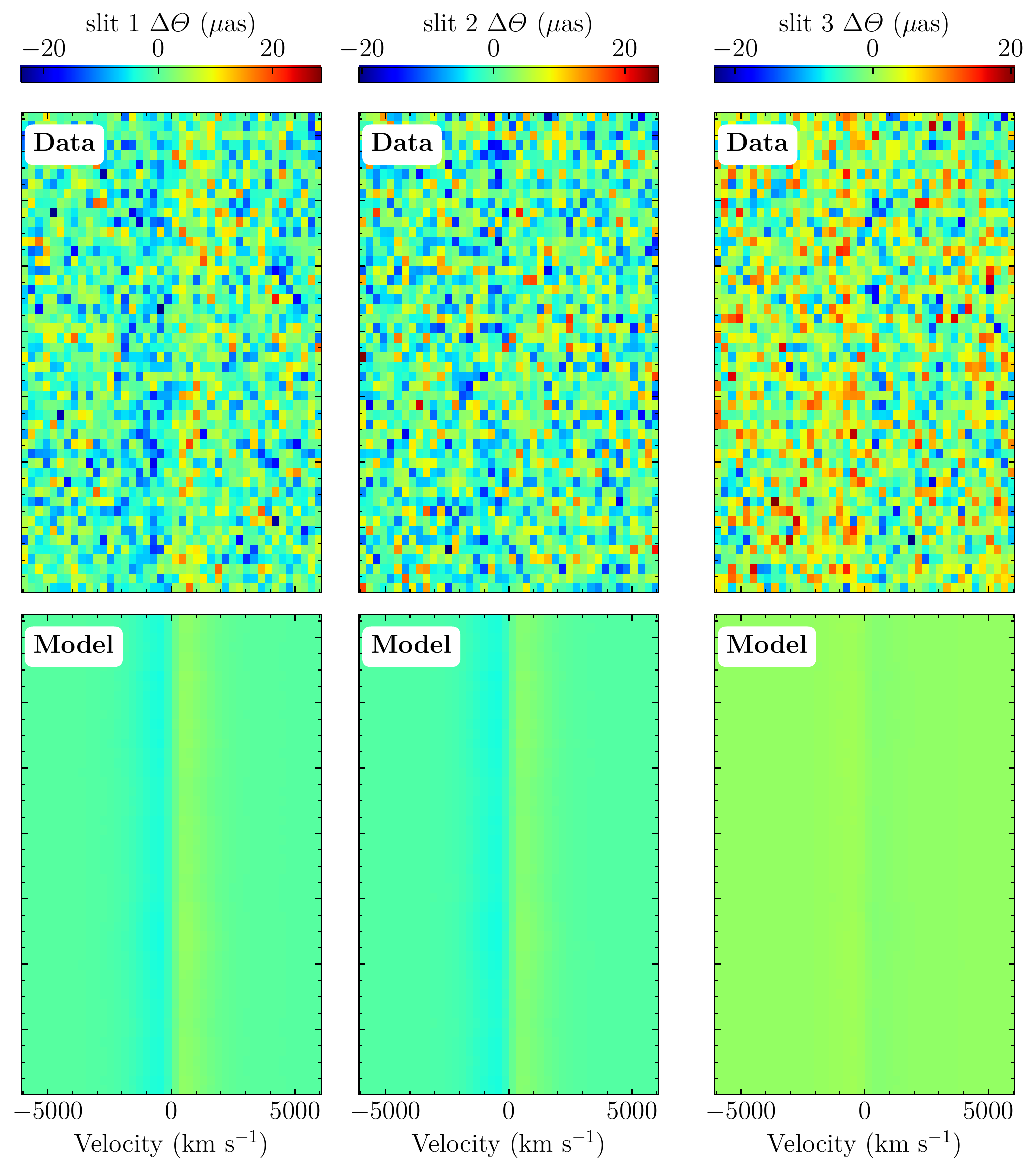}
\caption{The simulated photocenters and the recovery for the case of the input photocenter errors of 5 $\mu$as, which are comparable to the  maximum shift of the true photocenters illustrated in Figure~\ref{fig_slit}.}
\label{fig_sim_sarm_data_error}
\end{figure}

\subsubsection{Dependence on the Sampling Rate}
We randomly discard a fraction of epochs of the line profile and photocenter data generated in the preceding section and obtain a set of new data with line epochs ranging from 2 to 50. {\it Note that the sampling of the continuum light curve remains unchanged.} We rerun our package and summarize the recovered parameter values and uncertainties for different numbers of epochs as shown in the left panel of Figure~\ref{fig_para_values}. As expected, the uncertainties gradually decrease as the epochs increase. It is worth mentioning that even in the case of two epochs, we can still reasonably constrain the BLR size, black hole mass, and angular-size distance, albeit with relatively large uncertainties.
Figure~\ref{fig_comp_epoch} compares the posterior distributions of the model parameters for the cases of two epochs and nine epochs, from which, as expected, we can find stronger degeneracy between the BLR size and angular-size distance in the former case. The reasons that the BLR model parameters are reasonably constrained for a few line epochs are twofold. (1) The high fidelity of the continuum light curve ensures a meaningful detection of time delays. (2) The same BLR model is used for input and outputs so that there are no systematic errors arising from a possible BLR model mismatch. It is worth further investigating the second point in a future work.

\subsubsection{Dependence on the Measurement Errors of Photocenters}
\begin{figure*}[t!]
\centering
\includegraphics[width=0.7\textwidth]{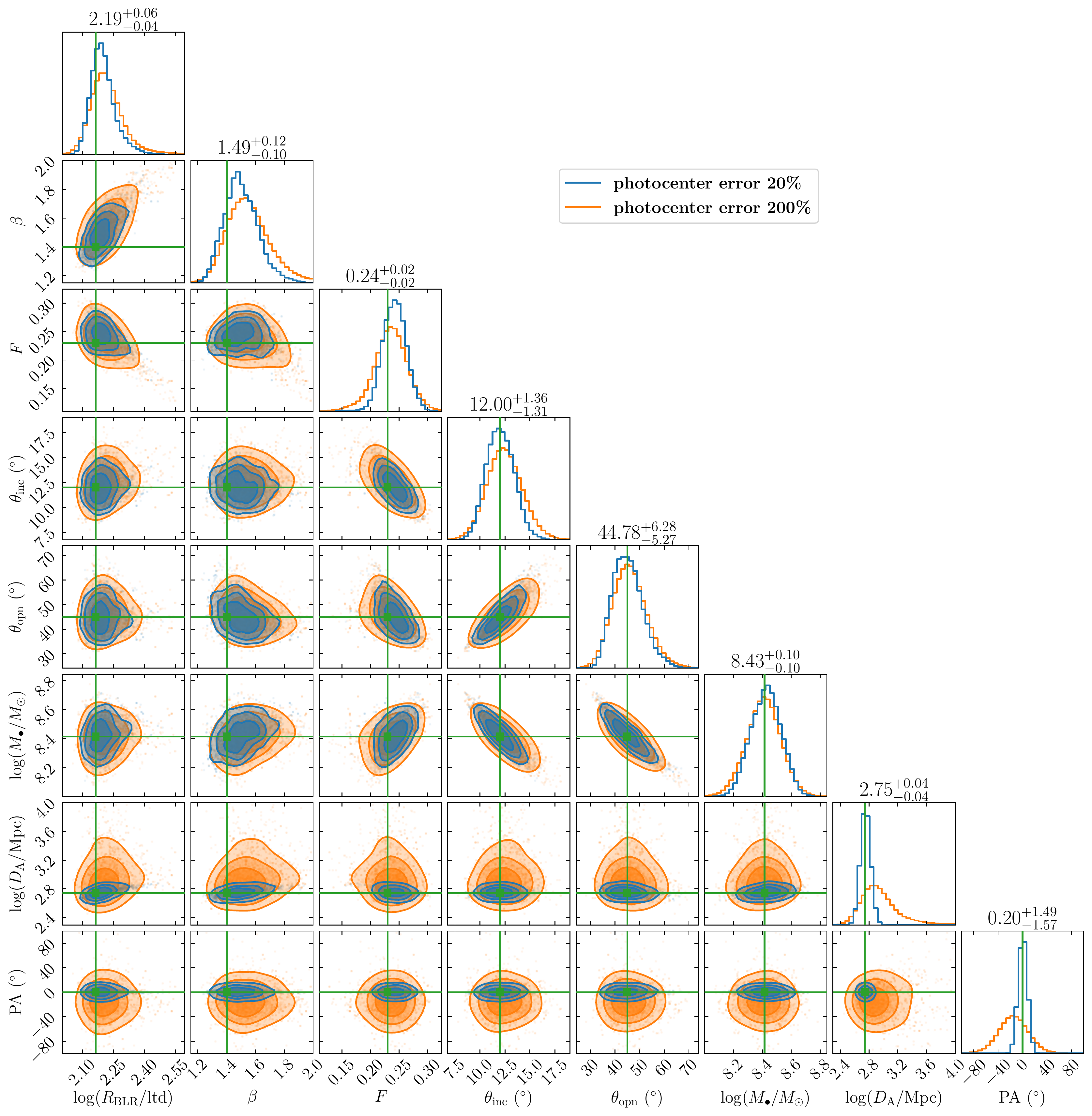}
\caption{A comparison of the inferred posterior distributions of the model parameters for the relative photocenter errors of 20\% (blue) and 200\% (yellow).
Here, the relative photocenter error means the photocenter error relative to the maximum shift of the true photocenters ($\sim$5 $\mu$as).
Green lines represent the input values. The contours are at the 1$\sigma$, 1.5$\sigma$, and 2$\sigma$ levels. The numbers above each diagonal panel mark the media value and 68.3\% confidence intervals from the case of the photocenter error of 20\%.}
\label{fig_comp_error}
\end{figure*}

Considering that the errors of the continuum and line profile are reasonable with existing telescopes, we only concentrate on the errors of the photocenters.
We change the input errors of the photocenters from 0.1 to 10 $\mu$as (while keeping the errors of the line profiles unchanged) and generate a set of new simulated data. The right panel of Figure~\ref{fig_para_values} illustrates the recovered parameter values and uncertainties for different input errors. Figure~\ref{fig_sim_sarm_data_error} shows the simulated data
and the recovery for the case of input errors of 5 $\mu$as, which are comparable to the maximum shift of the true photocenters (see Figure~\ref{fig_slit}). Because the errors of the line profiles are not changed, large input photocenter errors
mean that the likelihood from intensity RM becomes dominant over that from spectroastrometric RM. As a result, the BLR model parameters can still be well constrained, except for angular-size distance
and P.A., of which the uncertainties increase rapidly as the input photocenter errors are comparable  to or larger than 5 $\mu$as. In Figure~\ref{fig_comp_error},
we compare the posterior distributions of the model parameters between the input photocenter errors of 1 and 10 $\mu$as. In the latter case, the photocenter error is 2 times the maximum shift of the true photocenters (5 $\mu$as). The angular distance and P.A. have broad distributions, but from which reasonable inferences can be made.

\begin{figure*}
\centering
\includegraphics[height=0.4\textwidth]{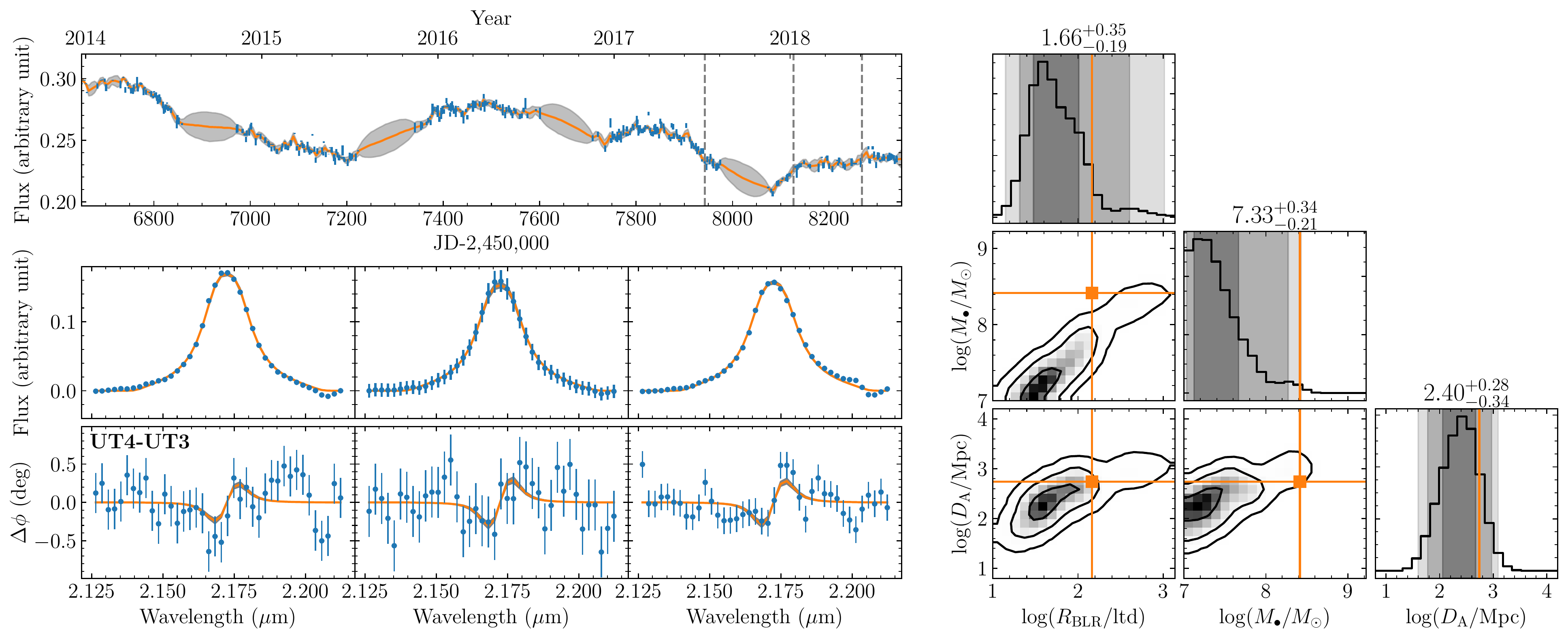}
\caption{Fits to the spectroastrometric data of 3C 273. (Left) The continuum light curve (top), Pa$\alpha$ line profiles (middle), and differential phase curves
of one baseline UT4-UT1 (bottom). The yellow solid line with shaded bands shows the model reconstructions. In the top panel, three vertical dashed lines mark the epochs when the GRAVITY observations were taken. (Right) The posterior distributions of BLR size ($R_{\rm BLR}$), black hole mass ($M_\bullet$), and angular-size distance ($D_{\rm A}$). The yellow solid lines represent the best inferences to $R_{\rm BLR}$ and $M_\bullet$ from \cite{Gravity2018} and the corresponding angular-size distance of 550 Mpc at the redshift of 3C~273 under the standard $\Lambda$CDM cosmology, respectively. The contours are at the 1$\sigma$, 2$\sigma$, and $3\sigma$ levels.}
\label{fig_fit_3c273}
\end{figure*}

\begin{figure*}
\centering
\includegraphics[width=0.7\textwidth]{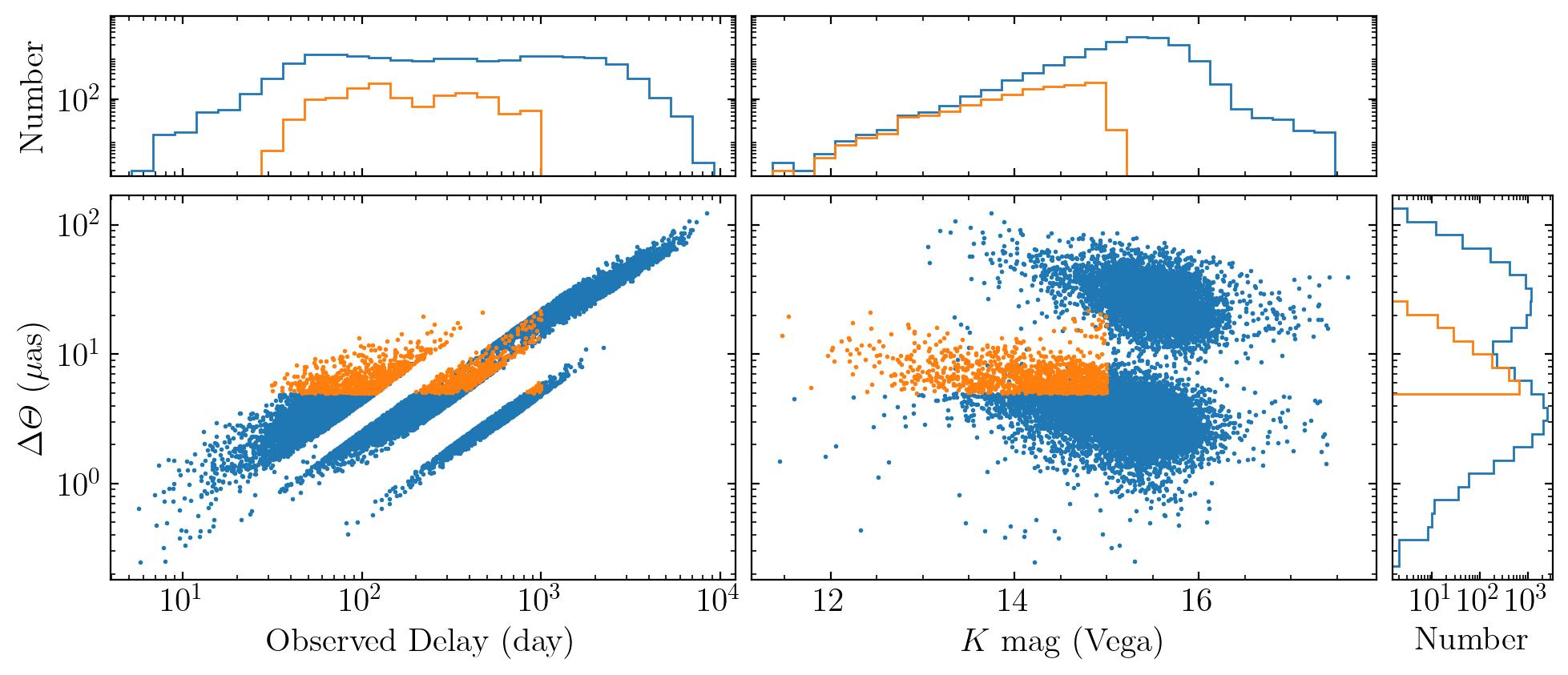}
\caption{A sample of AGNs from the SDSS DR14 with the estimated maximum shifts of the photocenters, observed time delays, and available $K$-band magnitude (Vega). The yellow points
represent the maximum shift of photocenters larger than 5 $\mu$as, time delays shorter than 10$^3$ days, and $K$-band magnitude brighter than 15 mag. See the text for the adopted ratios of $F_{\rm BLR}/F'_{c}$.}
\label{fig_sample}
\end{figure*}

\section{Application to 3C 273}\label{sec_3c273}
The preceding section illustrates the validity of our Bayesian approach. We now apply it to the spectroastrometric data of the Pa$\alpha$ line in 3C 273 observed by the GRAVITY/VLTI instrument (\citealt{Gravity2017}). The instrument coherently combines the light from the four 8 m telescopes and measures the interferometric phases on each of the six baselines.  The observations were taken on eight nights between 2017 July and 2018 May and those exposures of adjacent nights were combined to improve the S/Ns. For the details of the observations, refer to \cite{Gravity2018}. The GRAVITY Collaboration kindly shared the data, which originally consisted of four epochs (see \citealt{Gravity2018}).  However, since spectroastrometric RM analysis needs both the epoch-dependent line profiles and SA of the Pa$\alpha$, it turned out that there was one epoch without observing a calibration star, so the obtained Pa$\alpha$ line profile of that epoch cannot be used for the present purpose. As a result, there are three epochs of usable data, as shown in Figure~\ref{fig_fit_3c273} and Appendix~\ref{app_3c273}.

In running the analysis code, we use the same parameter priors as listed in Table~\ref{tab_param} and set the redshift of 3C 273 to be $z=0.158$. Because the GRAVITY observations used the adaptive optics systems, there was difficulty in calibrating the absolute fluxes and the measured Pa$\alpha$ line fluxes were normalized by the underlying continuum. We assume that the continuum variations underlying the Pa$\alpha$ follow those of the $V$-band light curve (\citealt{Li2020}) but with a time delay of about 450 days (\citealt{Sobrino2020}). As such, we determine the continuum fluxes underlying the Pa$\alpha$ line by interpolating the the $V$-band light curve after correcting the time delay and thereby obtaining the absolute fluxes of the Pa$\alpha$ line, for which the uncertainties from the interpolation are also included. We finally use the $V$-band light curve as a proxy for the driving continuum and take into account the delay of 450 days in the spectroastrometric RM analysis.

In the left panels of Figure~\ref{fig_fit_3c273}, we plot the reconstructions to the continuum light curve, Pa$\alpha$ line profiles, and the differential phase curves of one baseline UT4-UT3. The full fits to the differential phase curves of all six baselines are shown in Appendix~\ref{app_3c273}. The right panels of Figure~\ref{fig_fit_3c273} plot the obtained posterior distributions of the BLR size ($R_{\rm BLR}$), black hole mass ($M_\bullet$), and angular-size distance ($D_{\rm A}$). There appears strong degeneracy among the three parameters, implying that they are not well constrained and have large uncertainties. This is not surprising considering the noisy differential phase curves and only three epochs of data. While $M_\bullet$ tends to approach the lower prior limit of $10^7M_\odot$, $R_{\rm BLR}$ and $D_{\rm A}$ are peaked within their prior ranges, which are set to ($10$, $1300$) light-days and ($10$, $10^4$) Mpc, respectively. For the sake of comparison, we superimpose the best inferences on the BLR size and black hole mass from \cite{Gravity2018}. Our results are comparable to these inferences within a confidence level of $\leqslant$3$\sigma$. Note that the analysis of \cite{Gravity2018} did not include the spectroastrometric RM, therefore, they needed to preset the angular-size distance (550 Mpc). Remarkably, our obtained angular-size distance $\log(D_{\rm A}/{\rm Mpc})=2.40_{-0.34}^{+0.28}$ is marginally consistent within uncertainties with this fiducial value.

{\it We stress that our present application to 3C~273 is somehow tentative in light of the data quality.} In particular, the obtained black hole mass reaches its lower prior limit, indicating that to some extent, the adopted priors play a role in constraining the model parameters.  Nevertheless, the results are still enlightening. In the future, more epochs of observations with improved data quality for 3C~273 would be highly worthwhile to reinforce our present analysis and most importantly, to reliably measure the black hole mass and geometric distance of 3C~273. This is practically feasible considering the forthcoming upgraded instrument GRAVITY+ (see \citealt{Gravity2022}).

\section{Discussion}\label{sec_dis}
\subsection{Practical Concerns with Spectroastrometric RM}\label{sec_accuracy}
The theoretical statistical $1\sigma$ astrometric accuracy depends on the full width at half maximum (FWHM) of the point spread function (PSF) of the telescope and the number of photons collected per spectral bin $dN_{\rm ph}/d\nu$ as (\citealt{Stern2015, Bosco2021}), which can be estimated by
\begin{eqnarray}
\sigma_s &=& \frac{\rm FWHM_{\rm PSF}}{2.35 \left(\frac{dN_{\rm ph}}{d\nu}\right)^{1/2}} \nonumber\\
&=& 21.3~\mu{\rm as}\left(\frac{\rm FWHM_{\rm PSF}}{50~\rm mas}\right)
\left(\frac{1}{10^6}\frac{dN_{\rm ph}}{d\nu}\right)^{1/2}.
\end{eqnarray}
For the GRAVITY interferometer instrument, the baselines are on the order of 100 m, corresponding to $\sigma_s\sim3~\mu$as at the $K$ band. For the 30m class single-aperture telescopes, the expected accuracy increases up to $\sigma_s\sim10~\mu$as. Those estimates are based on the instruments being in ideal conditions (excluding all other sources of errors) and a presumption of $10^6$ photons per spectral bin. However, in practice, there are a variety of subtle statistical and systematic error sources that limit the achievable astrometric accuracy, such as atmospheric differential tilt jitter, anisoplanatism of AO systems, and atmospheric differential chromatic refraction, etc (e.g., \citealt{Cameron2009, VanBelle2009, Trippe2010}; \citealt{Rodeghiero2021}). All these error sources can dilute the final achievable astrometric accuracy at different levels. An advantage here is that spectroastrometric RM only requires differential astrometry across small spatial scales so that some of the issues related to absolute astrometry are no longer important.

By taking into account all practical sources of errors, previous studies have shown that the forthcoming 30m class single-aperture telescopes are likely to achieve a resolution at the level of several tens to approximately 100 microarcseconds at the $K$ band (e.g., \citealt{Trippe2010, Stern2015, Bosco2021,Rodeghiero2021}).
The observations with the GRAVITY instrument have demonstrated that it can achieve an angular resolution down to $\sim$10 $\mu$as at the $K$ band given sufficiently bright targets as well as adequate exposure times (usually several hours; \citealt{Gravity2017, Gravity2018, Gravity2022}). These factors, together with our application to 3C~273 in Section~\ref{sec_3c273}, indicate that it is viable to conduct spectroastrometric RM experiments on bright AGNs. Such experiments would span several months or years, depending on the BLR sizes, so as to capture the reverberation signals. Simulation tests in Section~\ref{sec_tests} imply that several epochs of SA observations already yield meaningful constraints on BLR model parameters. Nevertheless, we bear in mind the difficulties in performing SA observations. Currently, there are only a few AGNs with SA observation results published (\citealt{Gravity2018, Gravity2020, Gravity2021, Bosco2021}).
This situation might change with the upgraded GRAVITY+ and next-generation 30 m class telescopes.

\begin{figure*}[t!]
\centering
\includegraphics[width=0.9\textwidth]{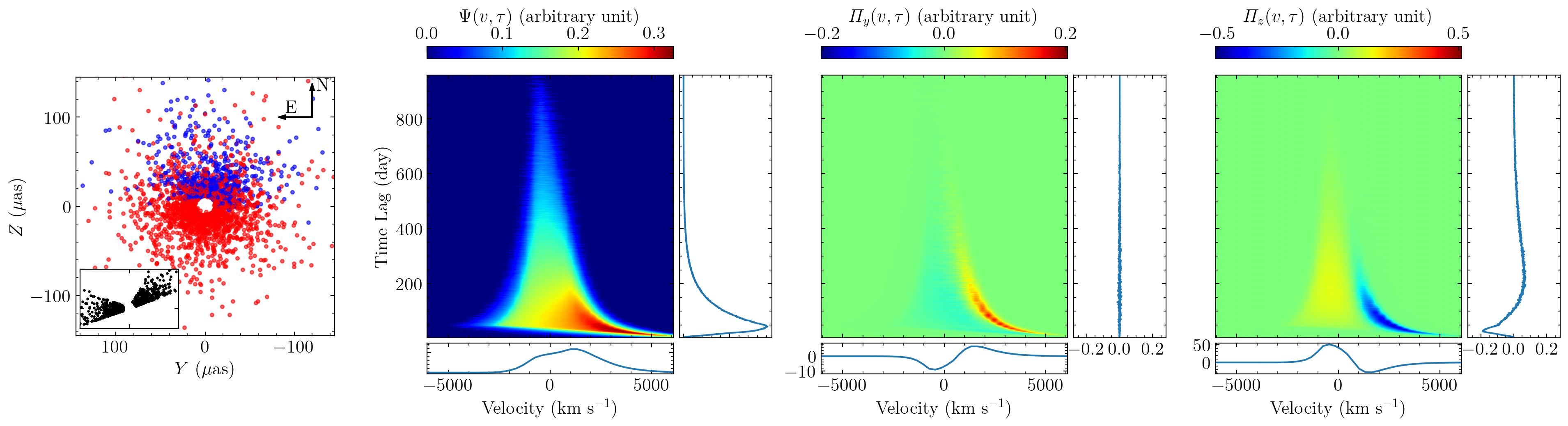}
\includegraphics[width=0.9\textwidth]{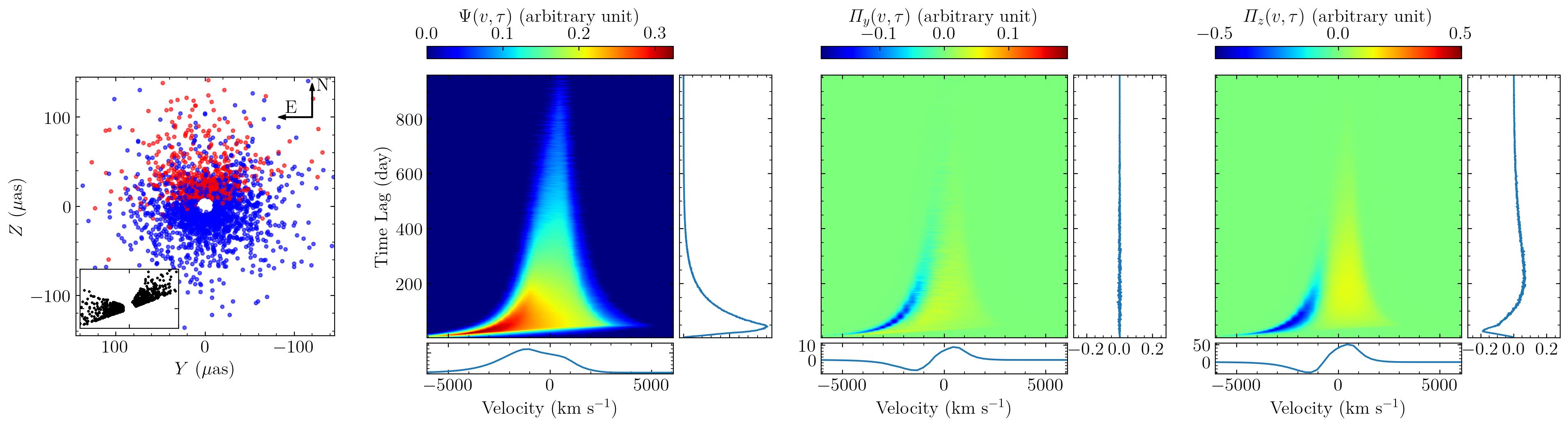}
\includegraphics[width=0.9\textwidth]{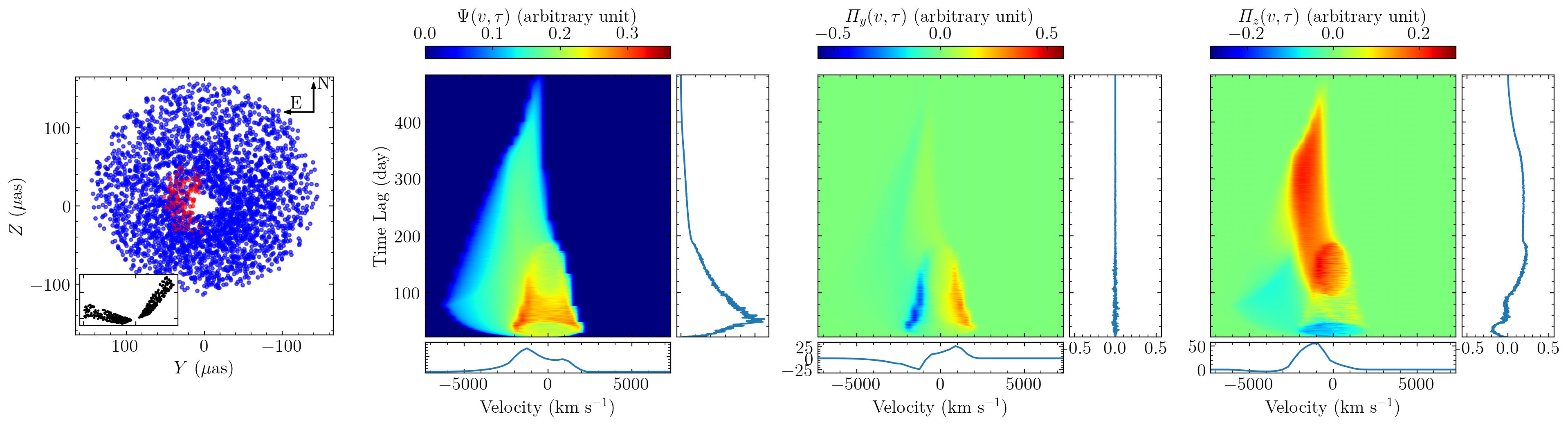}
\caption{Transfer functions for different BLR dynamical models. The top and middle panels illustrate the inflow and outflow models (\citealt{Pancoast2014}), respectively. The bottom panel illustrates the disk wind model (\citealt{Higginbottom2013}). The leftmost panels show the distributions of BLR clouds projected on the sky and the insets show the edge-on cutaway ($Z$-$X$) view of BLR clouds. The observed inclination is 20$^\circ$.}
\label{fig_model}
\end{figure*}

\subsection{Angular Sizes of BLRs from the Sloan Digital Sky Survey}
In this section, we estimate angular-size distributions of the quasar catalog from the 14th data release (DR) of the Sloan Digital Sky Survey (SDSS; \citealt{Paris2018}) to illustrate that there are available potential candidate targets for future spectroastrometric RM experiments.
\cite{Rakshit2020} performed detailed spectral measurements for the quasars with the continuum S/N $>$3 per pixel through multicomponent spectral decompositions, which included host-galaxy subtraction so that the central AGN luminosities could be obtained directly. We convert the luminosities to BLR sizes (or time delays) using the BLR size-luminosity relationship of \cite{Bentz2013} (see also \citealt{Du2019}). This relationship relies on the 5100~{\AA} luminosity and only corresponds to the H$\beta$ BLR size. For simplicity, we assume that all broad emission lines appropriate for infrared SA observations have the same BLR sizes as the H$\beta$ line. For those high-redshift AGNs ($z>0.8$) without available 5100~{\AA} luminosities, we make estimates from the given 3000 or 1350~{\AA} luminosities using the bolometric correction factors in \cite{Richards2006}. The observed time delays are given by multiplying the BLR sizes with the redshift factor ($1+z$).

Equation~(\ref{eqn_dphoton}) demonstrates that the observed photocenters scale with the normalized line fluxes relative to the underlying continuum flux, namely, the flux ratio $F_{\rm BLR}(v, t)/F'_{c}(t)$. The peak flux ratio generally differs among AGN emission lines, e.g.,  the typical value is $\sim$0.06 for the Br$\gamma$ line,  $\sim$0.3 for Pa$\beta$, 0.6 for Pa$\alpha$, 0.12 for Pa$\gamma$, and $>$1 for H$\alpha$ (\citealt{Landt2008, Rakshit2015}). Using the suitable emission line for the $K$ band at different redshifts, we simply assume a peak flux ratio of 0.06 for $z<0.08$, 0.6 for $0.08\leqslant z<0.4$, 0.3 for $0.4\leqslant z<0.87$, 0.12 for $0.87\leqslant z<1.26$, and 1.0 for $z\geqslant1.26$.
All the calculations assume a $\Lambda$CDM cosmology with $H_0=70~{\rm km~s^{-1}~Mpc^{-1}}$, $\Omega_{\rm M}=0.3$, and $\Omega_{\rm \Lambda}=0.7$.

Figure~\ref{fig_sample} shows the distributions of the maximum shifts of the photocenters, observed time delays, and $K$-band (Vega) magnitudes. Here, the $K$-band magnitudes are compiled by \cite{Paris2018} from the Two Micron All Sky Survey data. The yellow points in Figure~\ref{fig_sample} represent those AGNs with the maximum shift of photocenters $>5$ $\mu$as, observed time delays $<1000$ days, as well as magnitudes $K<15$, which can serve as candidates for future spectroastrometric RM experiments within a reasonable time span.

\subsection{Constraining the BLR Models}
The present dynamical modeling approach requires presuming a BLR model. A naturally arising issue is how the results depend on the presumed BLR model.
As in our previous works of \cite{Li2018} and \cite{Li2022}, an appropriate way to address this issue is testing a suite of BLR models and singling out the most probable one using statistical methods (such as Bayesian model selection). In some cases, there exist other independent observational measurements/constraints, with which we can further prove the validity of the selected model.

For the sake of illustration, in Figure~\ref{fig_model}, we showcase the transfer functions for three exemplary BLR models: the inflow and outflow prescription of \cite{Pancoast2014}, and the disk wind model of \cite{Higginbottom2013}. We stress that these three phenomenological BLR models are still restrictive and do not represent the whole story of BLRs, considering the complicated BLR kinematics (e.g., \citealt{Baskin2014,Czerny2017, Wang2017}). Notwithstanding, as can be seen, there are distinctive patterns in both the intensity and spectroastrometric transfer functions among the three models. We expect that appropriate analysis with spectroastrometric RM data can better constrain and prove the validity of different BLR models, compared to the traditional intensity RM.

\subsection{Comparison with the Previous Joint Analysis of SA and RM}
As mentioned above, \cite{Wang2020} proposed jointly analyzing SA and intensity RM (hereafter SARM) data to probe the BLR geometry and kinematics and measure the geometric distance.  They made the first such application to 3C 273, which had been observed with infrared SA of the Pa$\alpha$ (\citealt{Gravity2018}) line and optical intensity RM of the H$\beta$ line (\citealt{Zhang2019}). Currently, SA is only feasible in infrared whereas almost all intensity RM campaigns are undertaken in the optical, therefore, their respectively observed emission lines are indeed different.
As a result, in such joint SARM analysis, one needs to presume that the two different lines share the same BLR. However, \cite{Li2022} showed that the profiles of the H$\beta$ and Pa$\alpha$ lines in 3C~273 differ regarding both widths and shapes. Such differences appear to be common in AGNs (\citealt{Landt2008}; see also Figure 2 of \citealt{Li2022}), implying that the BLRs must be different in some respects. \cite{Li2022} proposed using velocity-resolved intensity RM data and treating the respective BLRs separately, but leting them share only the inclination angle and SMBH mass. As such, the differences between the two lines are naturally taken into account. The disadvantage of this approach is a longer parameter list, which likely results in relatively large parameter uncertainties.

The spectroastrometric RM circumvents the above issue since it observes the intensity and SA of the same line; therefore, it gets rid of possible systematic errors arising from different lines in previous joint SARM analyses. Currently, the challenges of SA observations still restrict the targets of spectroastrometric RM to a few bright AGNs; however, as mentioned above, the situation might significantly change with the forthcoming upgraded instrument GRAVITY+, which is planned to make performance improvements in several key respects, such as the interferometric fringe tracking and sensitivity magnitude (\citealt{Gravity2022}).

\section{Conclusions}\label{sec_conclusion}
We propose that spectroastrometric signals of BLRs in AGNs reverberate to the continuum variations and the responses of different BLR parts show different time delays as a result of spatial distributions of BLR gas (see also \citealt{Shen2012}). Considering that SA resolves the BLR structure perpendicular to the LOS, spectroastrometric RM, therefore, provides a new diagnostic for BLR geometry and kinematics, complementary to the traditional intensity RM technique. We present the basic mathematical framework for spectroastrometric RM, which indeed can be regarded as a deconvolution problem (see Equation~\ref{eqn_sarm_all}) so that a variety of well-established mathematical methods are applicable. The underlying essence of spectroastrometric RM is to determine the intensity and spectroastrometric transfer functions, which encode the full information about the BLR. We derive analytical expressions for the case of an inclined ring-like BLR and also show that in extreme cases where the continuum has a pulsing variation or remains constant, there exist simple expressions. For a generic BLR with realistic parameter values, the spectroastrometric signals vary on a level of several to tens of microarcseconds, mainly depending on the BLR size, continuum variability, and cosmic distance.

We developed a forward Bayesian dynamical modeling approach to analyze spectroastrometric RM data and infer BLR properties, in which the posterior probability is explored with the MCMC technique. We constructed a suite of simulation tests to demonstrate the validity of our approach and show the potential of spectroastrometric RM in resolving BLR geometry and kinematics and most importantly measuring the SMBH mass and angular-size distance. An application to the spectroastrometric data of 3C~273 yields tentative, but enlightening constraints on its BLR size, central SMBH mass, and angular-size distance, although there are large uncertainties (see Figure~\ref{fig_fit_3c273}). Despite the challenges remaining in SA observations, these results remarkably imply the feasibility of conducting pilot spectroastrometric RM experiments on nearby bright AGNs, in particular, considering the forthcoming upgraded GRAVITY+/VLTI and the panned next-generation 30 m class telescopes.

\section*{Acknowledgements}
We thank the referee for useful comments that improved the clarity of the manuscript.
We also thank the GRAVITY Collaboration for kindly sharing the GRAVITY observational data of 3C~273 through E. Sturm and J. Shangguan.
We acknowledge financial support from the National Key R\&D Program of China (2021YFA1600404), from the National Natural Science Foundation of China (NSFC; 11833008 and 11991054), from the CAS International Partnership Program (113111KYSB20200014), and from the China Manned Space Project (CMS-CSST-2021-A06).
Y.R.L. acknowledges financial support from NSFC through grant Nos. 11922304 and 12273041 and from the Youth Innovation Promotion Association CAS.

\software{\texttt{CDNest} (\citealt{Li2020cdnest}), \texttt{BRAINS} (\citealt{Li2018}).}

\appendix

\section{Definition of the CCF}\label{app_ccf}
The CCF between two time series, say, $X(t)$ and $Y(t)$, is defined to be (e.g., \citealt{Welsh1999})
\begin{equation}
{\rm CCF}(X, Y, \tau) = \frac{1}{\sigma(X)\sigma(Y)} E\left(\left[X(t) - E(X)\right]\left[Y(t+\tau) - E(Y)\right] \right),
\end{equation}
where $\sigma(X)$ represents the standard variation of $X(t)$ and $E(X)$ represents the expectation of $X$. It is easy to prove that
\begin{equation}
{\rm CCF}(aX+d, cY+d, \tau) = {\rm CCF}(X, Y, \tau),
\end{equation}
where $a$, $b$, $c$, and $d$ are time-independent coefficients. Given with three time series, say, $X(t)$, $Y(t)$, and $Z(t)$,
\begin{equation}
{\rm CCF}(X+Y, Z, \tau) = \frac{1}{\sigma(X)+\sigma(Y)}\left[ \sigma(X){\rm CCF}(X, Z, \tau) + \sigma(Y){\rm CCF}(Y, Z, \tau)\right].
\end{equation}

\section{Deriving the Equations for a Ring-like BLR}\label{app_ring}
In the $XYZ$ frame, the LOS is $\bm{n}=(1, 0, 0)$ and a point $P$ (see Figure~\ref{fig_sch})
in the ring has a coordinate of $\bm{R} = R(\sin i\cos\theta, \cos\theta, -\cos i\cos\theta)$ and
a velocity of $\bm{w}=V(-\sin i\sin\theta, \cos\theta, \cos i\sin\theta)$, where $R$ and $V$ are the
the radius and rotating velocity of the ring, respectively. As a result,
the corresponding time delay of the point $P$ with respect to the origin $O$ is
\begin{equation}
\tau = \frac{R - \bm{R}\cdot\bm{n}}{c} = \frac{R}{c}(1 - \cos\theta \sin i).
\end{equation}
The LOS velocity is
\begin{equation}
v = -\bm{w}\cdot \bm{n} = V\sin i \sin\theta,
\end{equation}
and the projected location in the observer's sky is
\begin{equation}
\bm{r}_\perp = \bm{R} - (\bm{R}\cdot\bm{n})\bm{n} = R(0, \sin\theta, -\cos i\cos\theta).
\end{equation}
The intensity transfer function is given by
\begin{eqnarray}
\Psi(v, \tau) = \epsilon \int_0^{2\pi} \delta\left[v - V\sin i\sin\theta\right]
\delta\left[\tau - \frac{R}{c}(1-\cos\theta\sin i)\right]Rd\theta =  \frac{c\epsilon}{\rho}\left[\delta\left(v - V \rho\right)
 + \delta\left(v + V \rho\right)\right],
\end{eqnarray}
where $ R(1 - \sin i)/c\leqslant \tau \leqslant R(1+\sin i)/c$, otherwise $\Psi(v, \tau) = 0$, and
\begin{equation}
\rho = \sqrt{\sin^2i - (1 - c\tau/R)^2}.
\end{equation}
The spectroastrometric transfer function has a zero $x$-component, and its $y$- and $z$-components are given by
\begin{eqnarray}
\varPi_y(v, \tau)& = &\epsilon\int_0^{2\pi}R\sin\theta \delta\left[v - V\sin i\sin\theta\right]\delta\left[\tau - \frac{R}{c}(1-\cos\theta\sin i)\right]Rd\theta
=\frac{c\epsilon R}{\sin i }\left[\delta\left(v - V \rho \right) - \delta\left(v + V \rho\right)\right],
\end{eqnarray}
and
\begin{eqnarray}
 \varPi_z(v, \tau)& = &-\epsilon\int_0^{2\pi}R\cos i\cos\theta \delta\left[v - V\sin i\sin\theta\right]\delta\left[\tau - \frac{R}{c}(1-\cos\theta\sin i)\right]Rd\theta\\
& = & -\frac{\cos i}{\sin i}\frac{c\epsilon(R-c\tau)}{\rho} \left[ \delta\left(v - V \rho\right) + \delta\left(v + V \rho\right)\right].
\end{eqnarray}
The velocity integrals of the above transfer functions are
\begin{equation}
\tilde\Psi(\tau) = \frac{2c\epsilon}{\rho},
\end{equation}
and
\begin{equation}
\tilde\varPi_y(\tau) = 0, ~~~\varPi_z(\tau) = -\frac{\cos i}{\sin i}\frac{2c\epsilon(R-c\tau)}{\rho}.
\end{equation}
The delay integrals of the above transfer functions are
\begin{equation}
\hat\Psi(v) = \frac{2R\epsilon}{V\sqrt{\sin^2i-v^2/V^2}},
\end{equation}
and
\begin{equation}
\hat\varPi_y(v) = \frac{2R^2 \epsilon }{V^2\sin i}\frac{v}{\sqrt{\sin^2i-v^2/V^2}},~~~\hat\varPi_z(v) = 0.
\end{equation}

\section{Simulating Light Curves Using the DRW Model}\label{app_lc}
We generate mock light curves using the DRW model as follows. Given with the covariance matrix $\bm{C}$ of a DRW model, its Cholesky decomposition is written as $\bm{C=MM^T}$, where
$\bm{M}$ is a lower triangular matrix (\citealt[Chapter 2.9]{Press1992}). A mock light curve is obtained with $\bm{u=Mr}$, where $\bm{r}$ is a series of
Gaussian random numbers with a zero mean and unity deviation. It is easy to show that such a light curve $\bm{u}$ has a covariance matrix of
$\langle \bm{uu^T}\rangle=\bm{M}\langle \bm{rr^T} \rangle\bm{M^T}=\bm{C}$. Here, the covariance matrix of a DRW model is given by
\begin{equation}
C_{ij} = \sigma_{\rm d}^2\exp\left(-\frac{|t_i-t_j|}{\tau_{\rm d}}\right),
\end{equation}
where $t_i$ and $t_j$ are the times of $i$th and $j$th points of the light curve, respectively, and $\sigma_{\rm d}$ and $\tau_{\rm d}$ are parameters that represent the long-term standard variation and typical damping time scale of the DRW process.

\section{Calculating Spectroastrometric Signals Given a BLR Model}\label{app_sa}
After generating BLR clouds' velocities and positions according to the given BLR model (see, e.g., \citealt{Li2022}),
the transfer functions defined in Equations~(\ref{eqn_stf}) and (\ref{eqn_satf}) are calculated as
\begin{eqnarray}
\Psi(v, \tau) & = & \sum_i \delta(v-v_i) \delta(\tau-\tau_i) \epsilon_i,\\
\varPi_y(v, \tau) &=& \sum_i y_i\delta(v-v_i) \delta(\tau-\tau_i) \epsilon_i,\\
\varPi_z(v, \tau) &=& \sum_i z_i\delta(v-v_i) \delta(\tau-\tau_i) \epsilon_i,
\end{eqnarray}
where $\epsilon_i$, $v_i$,  $\tau_i$, $y_i$, and $z_i$ are the response coefficient, LOS velocity, time delay, and $y$ and $z$ coordinates
of the $i$ cloud, respectively. The flux and SA of the emission line are then calculated using Equations~(\ref{eqn_rm})
and (\ref{eqn_sarm}). For simplicity, we assume that all clouds have a uniform response coefficient $\epsilon_i$ and
the possible nonlinear response of the line emission to the continuum is neglected (\citealt{Li2013}).

\begin{figure}[th!]
\centering
\includegraphics[width=0.6\textwidth]{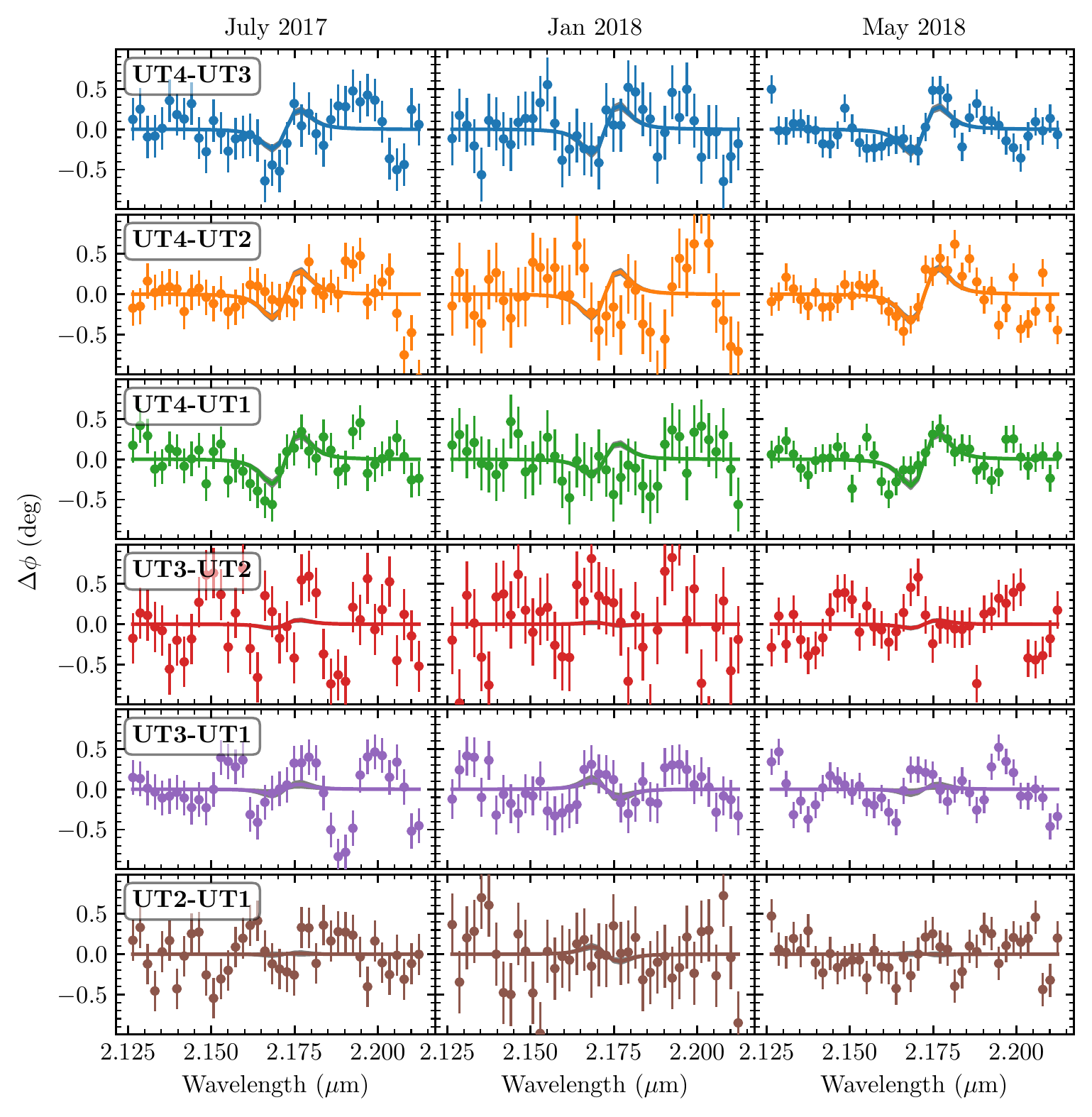}
\caption{Full fits to the three epochs of differential phase curves for the Pa$\alpha$ line in 3C 273 observed by the GRAVITY/VLTI (\citealt{Gravity2018}). The solid lines with gray shaded bands show reconstructions from model fits. The wavelengths are given in the observed frame.}
\label{fig_fit_3c273_full}
\end{figure}

\section{Full Fits to the Data of 3C~273}\label{app_3c273}
In Figure~\ref{fig_fit_3c273_full}, we show the full fits to the differential phase curves of the six baselines observed by the GRAVITY/VLTI
for the Pa$\alpha$ line of 3C~273 (\citealt{Gravity2018}). See Section~\ref{sec_3c273} for the details of the model fitting.

\end{document}